\documentclass[aps,prx,superscriptaddress,eprint,nofootinbib]{revtex4-2}
\usepackage{amsmath}
\usepackage{amsthm}
\usepackage{mathtools}
\usepackage{hyperref}
\usepackage{amssymb}
\usepackage{physics}
\usepackage{graphicx}
\usepackage{booktabs,amsfonts,dcolumn}
\usepackage{spacingtricks}
\graphicspath{ {images/} }
\usepackage[caption=false]{subfig}
\usepackage{bm}
\usepackage[dvipsnames]{xcolor}
\usepackage{float}

\usepackage{setspace} 

\usepackage[section]{placeins} 

\usepackage{comment}

\usepackage{empheq}

\usepackage{textcomp}
\usepackage{siunitx}

\usepackage{tikz}
\usetikzlibrary{positioning, calc}

\usepackage{adjustbox}

\usepackage{pdfpages}
\makeatletter
\AtBeginDocument{\let\LS@rot\@undefined}
\makeatother
\usepackage[portrait, margin=0.5in]{geometry}

\newcolumntype{d}[1]{D..{#1}}

\newcommand{\beq}{\begin{equation}}
	\newcommand{\eeq}{\end{equation}}

\begin{document}
	
	\title{Defect ground states for liquid crystals on cones and hyperbolic cones}
	\author{Farzan Vafa}
	\affiliation{Center of Mathematical Sciences and Applications, Harvard University, Cambridge, MA 02138, USA}
	\author{Grace H. Zhang}
	\affiliation{Department of Physics, Harvard University, Cambridge, MA 02138, USA}
	\author{David R. Nelson}
	\affiliation{Department of Physics, Harvard University, Cambridge, MA 02138, USA}
	\date{\today}
	
	\begin{abstract}
	
	This contribution is intended for Journal of Physics A: Mathematical and Theoretical Special issue on Non-equilibrium Dynamics in Complex Systems: Celebrating the Contributions of Uwe T\"{a}uber on his 60th Birthday. Cones with orientational order in the local tangent plane provide a soft matter analog of the Aharonov-Bohm effect. In this article, we first review recent work on two-dimensional liquid crystals with $p$-fold rotational symmetry ($p$-atics) on cones. By exploiting an analogy with electrostatics, we determine the ground state as a function of both the cone deficit angle and the liquid crystal symmetry $p$ for both free boundary conditions and tangential boundary conditions applied at the cone base. There is an effective topological charge $-\chi$ at the apex, where $2\pi\chi$ is the deficit angle. The ground states are in general frustrated due to parallel transport along the azimuthal direction on the cone. In the case of tangential boundary conditions, the ground state changes as a function of $\chi$, where the cone apex absorbs and emits quantized defect charges, with intricate dependence on both the deficit angle and $p$. We check our predictions numerically for a set of commensurate cone angles, whose surfaces can be polygonized as a perfect triangular or square mesh, and find excellent agreement. Cones with both free and tangential boundary conditions can also exhibit metastable states distinguished by quantized screening of the apex charge. We also present preliminary work on hyperbolic cones, where the Gaussian curvature singularity is negative at the apex. When free boundary conditions are applied at the base, the ground states are characterized by an effective topological charge at the apex, similarly to the case of conventional cones. However, when tangential boundary conditions are applied, decreasing the deficit angle (which is now negative) induces neutral defect pair nucleation at the apex followed by emission of a positive defect.
		
	\end{abstract}
	
	\maketitle
	
	\tableofcontents


\section{Introduction to liquid crystals}
	
Crystals, admired for their stunning appearance and versatile usage in semiconducting electronics and in jewelry and decorative items, are important features of everyday life. However, the fascinating realm of \emph{liquid} crystals sets them apart, as they are intermediate states of matter, between the solid and liquid phases. Discovered by Austrian botanist and chemist Friedrich Richard Reinitzer in 1888 and later named by physicist Otto Lehmann in 1904, liquid crystals have intrigued scientists for decades~\cite{dunmur2014soap}.

\begin{figure}[t]
	\centering
	\subfloat[$p=1$]
	{\includegraphics[width=.32\columnwidth]{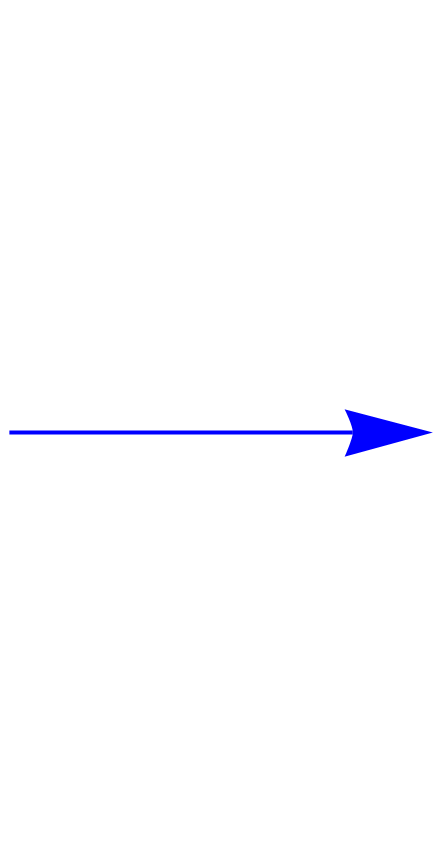}}
	\hfil
	\subfloat[$p=2$]
	{\includegraphics[width=.32\columnwidth]{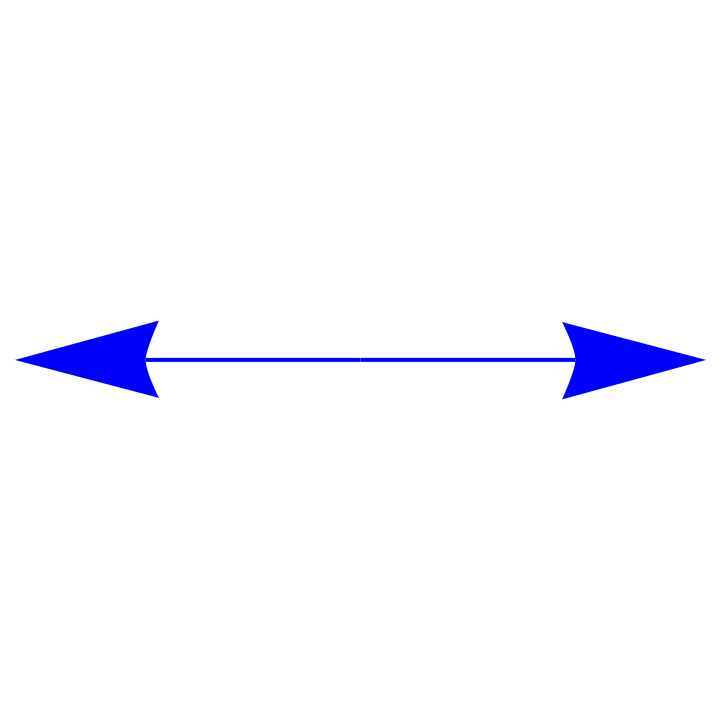}}
	\hfil
	\subfloat[$p=3$]
	{\includegraphics[width=.32\columnwidth]{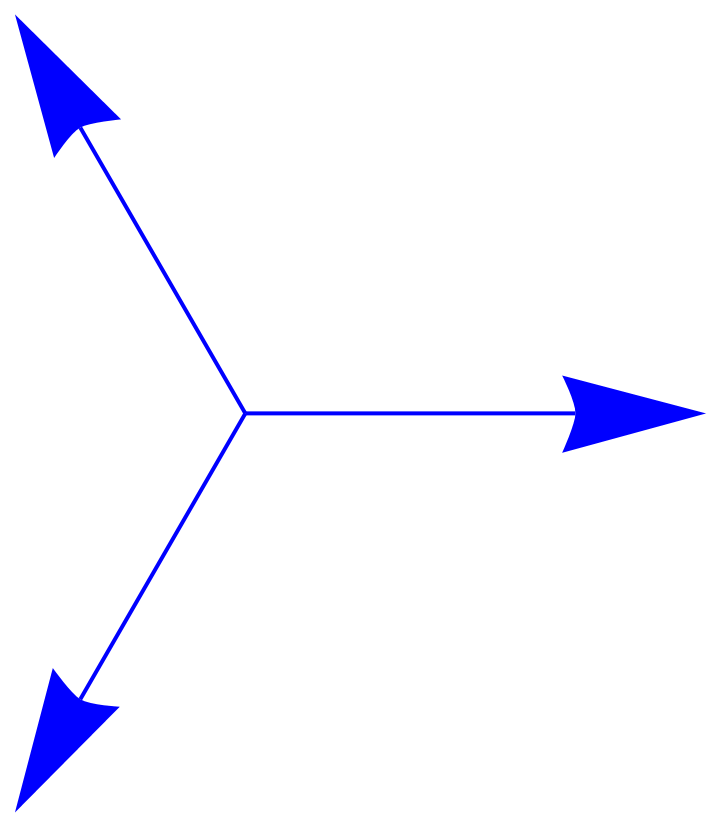}}
	\caption{Examples of $p$-atic units for $p = 1, 2, 3$. A polar ($p=1$) unit has one leg, a nematic ($p=2$) has two legs, and a triatic ($p=3$) unit has three legs. A $p$-atic is a collection of such units.}
	\label{fig:p-atics}
\end{figure}

What exactly are liquid crystals? Whereas conventional crystals possess both translational order and orientational order and liquids possess neither, liquid crystals possess orientational order but lack translational order~\cite{gennes1993the}. Nematic (the Greek root means ``threadlike'') liquid crystals flow like liquids, with however a broken directional symmetry that allows them to interact strongly with polarized light, important for many of their display applications. [In \emph{smectic} liquid crystals -- the Greek root means ``soaplike'' -- the molecules are arranged in layers, where the molecules within each layer are aligned.] For convenience, we call 2D nematic liquid crystals with $p$-fold rotational symmetry ``$p$-atic liquid crystals''~\cite{giomi2022hydrodynamic}, where $p$ denotes the type of orientational order. Nematic liquid crystals for example correspond to $p=2$. We can think of a $p$-atic more generally as a collection of units that each possess $p$ legs (see Fig.~\ref{fig:p-atics} for examples for $p = 1,2,3$). Applications of nematic and smectic liquid crystals range from electronic devices (LCDs)~\cite{yang2014fundamentals} to sunscreen~\cite{brinon1999percutaneous}; they're also being studied for biomedical applications~\cite{woltman2007liquid}, including targeted drug delivery~\cite{silvestrini2020advances}.

An equilibrium example of a $p=1$ liquid crystal in two dimensions is provided by tilted molecules: Langmuir-Blodgett monolayers at a liquid-air interface~\cite{kaganer1999structure}. Common examples of $p=1$ liquid-crystal-like vector order far from equilibrium in nature include bacteria with flagella migrating up a chemotactic gradient, a flock of birds, a school of fish, etc, where all of these energy-consuming organisms are moving together in the same direction on average, with centers of mass irregularly spaced, as in a liquid.

\begin{figure}[t]
	\centering
	\subfloat[$p=2$]
	{\includegraphics[height=.254\columnwidth]{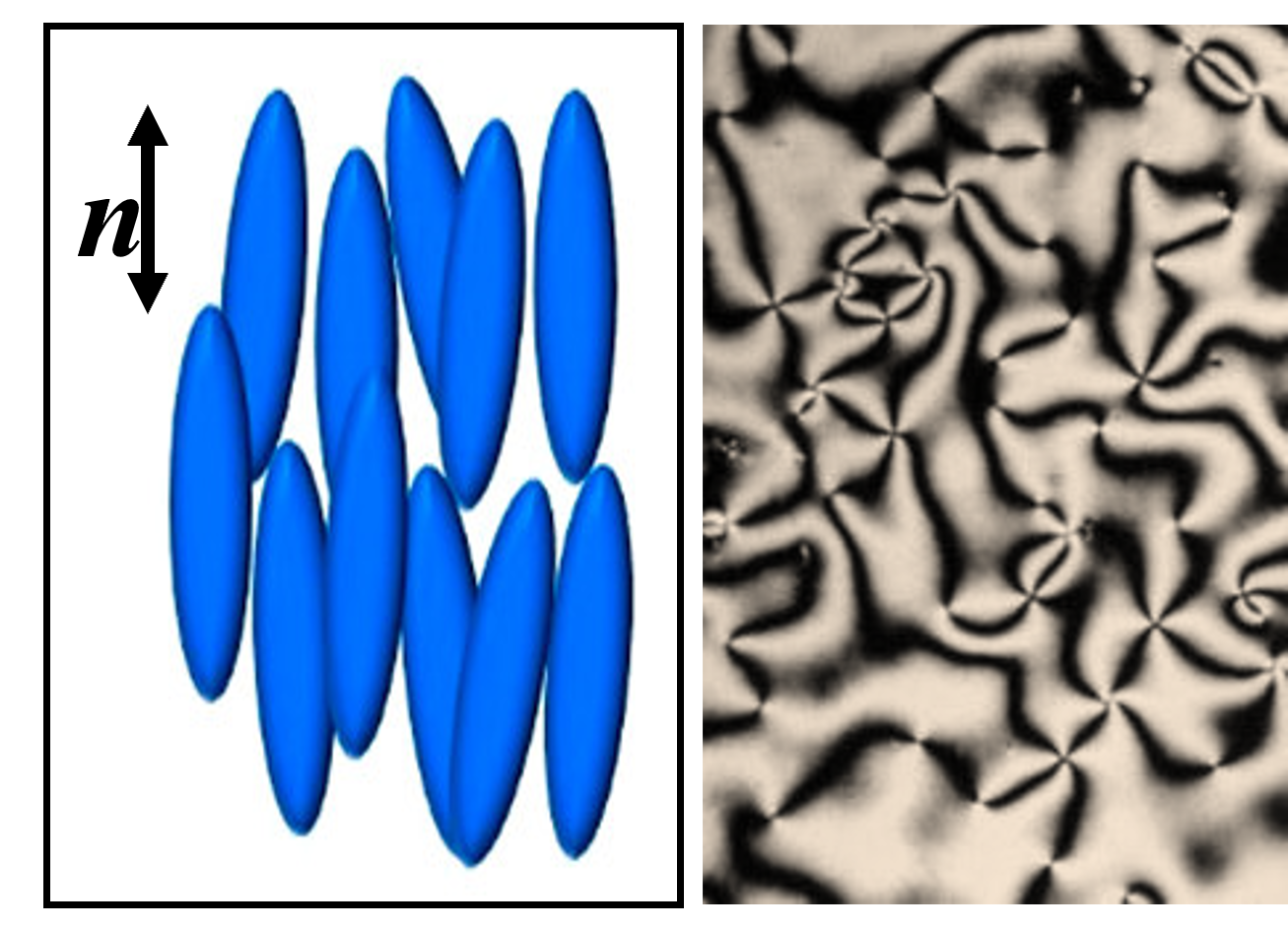}}
	\hfil
	\subfloat[$p=3$]
	{\includegraphics[height=.25\columnwidth]{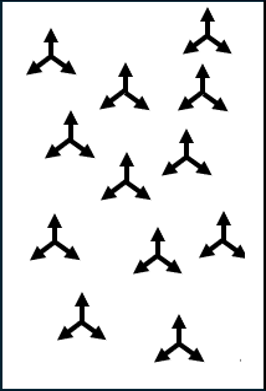}}
	\hfil
	\subfloat[$p=4$]
	{\includegraphics[height=.25\columnwidth]{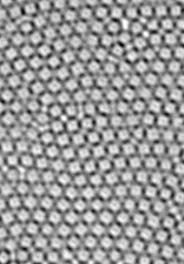}}
	\hfil
	\subfloat[$p=6$]
	{\includegraphics[height=.25\columnwidth]{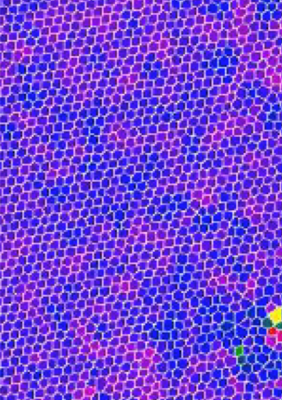}}
	\hfil
	\caption{Examples of $p$-atic liquid crystals in the laboratory for (a) $p=2$, (c) $p=4$, and (d) $p=6$, adapted from Refs.~\cite{dilisi2019introduction,loffler2018phase,thorneywork2017two}, respectively. (b) Hypothetical triatic liquid phase of matter in two dimensions, with extended order in the orientations of three-fold symmetric molecules whose arms point to the vertices on an equilateral triangle. (See Ref~\cite{zhao2012local} for a related material with two families of triatic molecules with staggered orientations.)}
	\label{fig:liquidCrystals}
\end{figure}

We have already discussed the case of $p=2$ nematic liquid crystals: the molecules, regarded as double-headed arrows (see Fig.~\ref{fig:p-atics}(b)) are all aligned in the same direction, but they're free to move around and rotate within that alignment. Note that although most rod-shaped liquid crystal molecules do not have a microscopic up-down symmetry~\cite{gennes1993the}, they behave statistically as double-headed arrows, with approximately equal numbers of molecules pointing parallel or antiparallel to the broken symmetry direction in a given region of alignment. Fig.~\ref{fig:liquidCrystals}(b) shows an usual collection of 3-legged molecular building block, which could lead to ``triatic'' order in a liquid, provided interactions in a liquid align the arms of the triad.

A more common type of orientational order arises in hexatic liquids ($p=6$), which exhibit six-fold orientational order. Potential examples in two dimensions include active cellular monolayers and colloidal suspensions, where the hexatic arises as a fourth equilibrium phase of matter, interposed between a crystal and a liquid~\cite{strandburg1988two}. See Fig.~\ref{fig:liquidCrystals} illustrating various types of $p$-atic liquid crystals.

\begin{figure}[t]
	\centering
	\subfloat[$+1$ defect]
	{\includegraphics[width=0.28\textwidth]{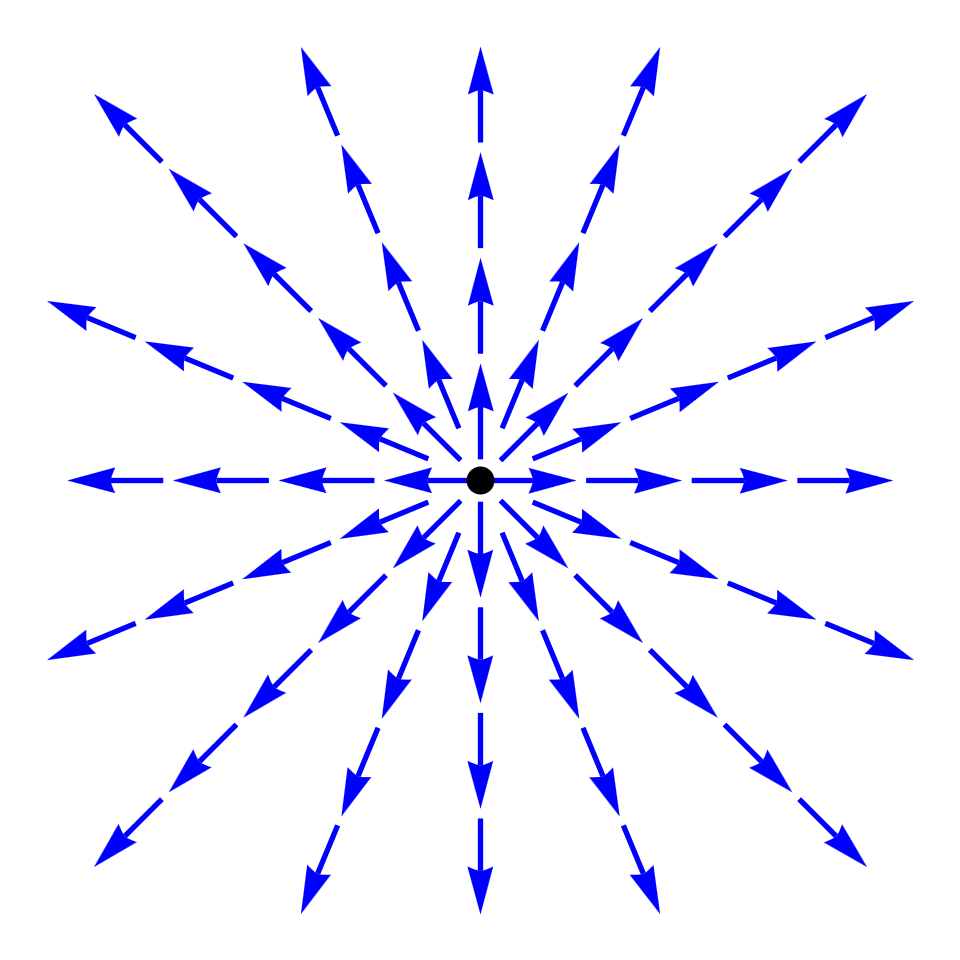}}
	\subfloat[$+1/2$ defect]
	{\includegraphics[width=0.28\textwidth]{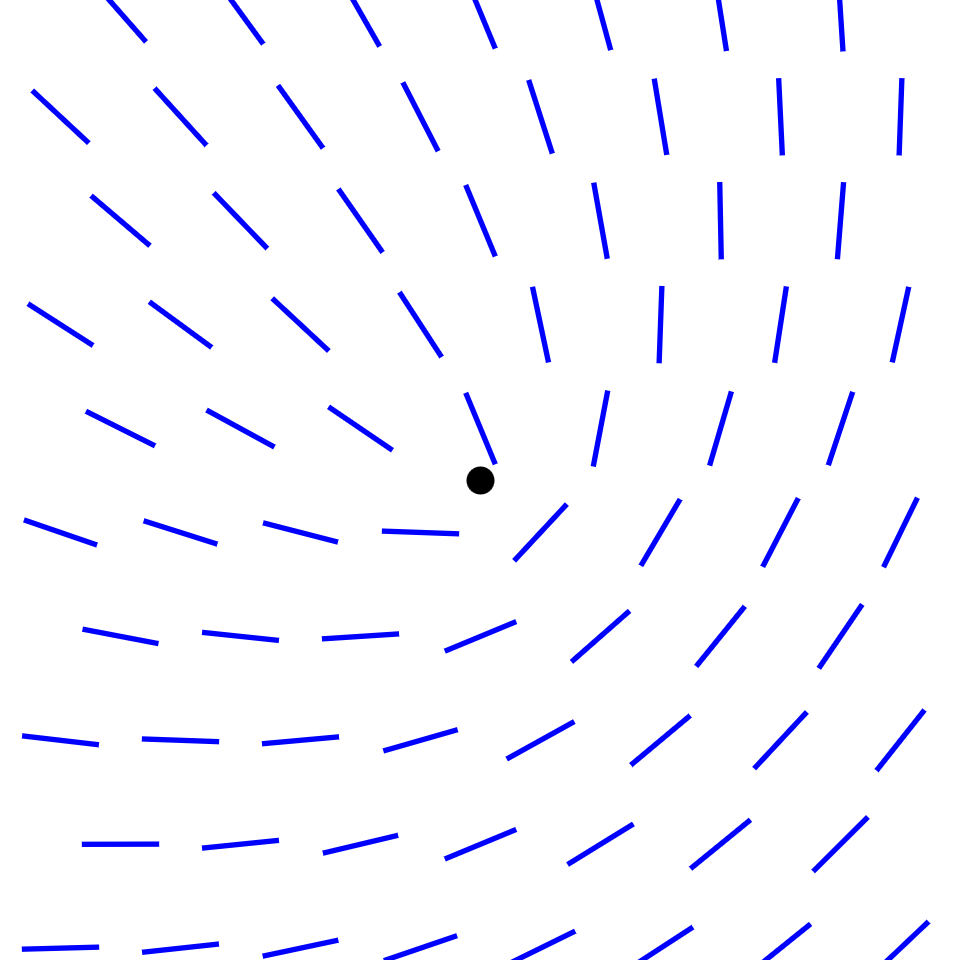}}
	\subfloat[$+1/3$ defect]
	{\includegraphics[width=0.28\textwidth]{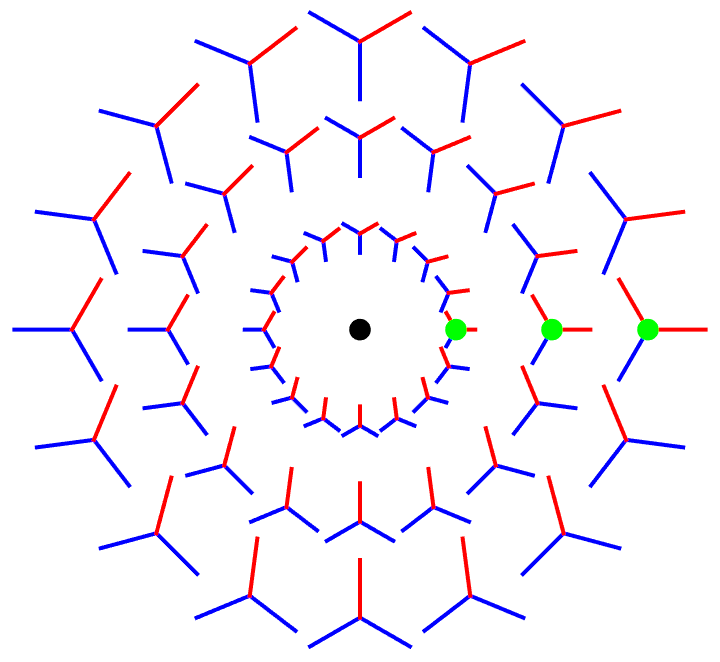}}\\
	\subfloat[$-1$ defect]
	{\includegraphics[width=0.28\textwidth]{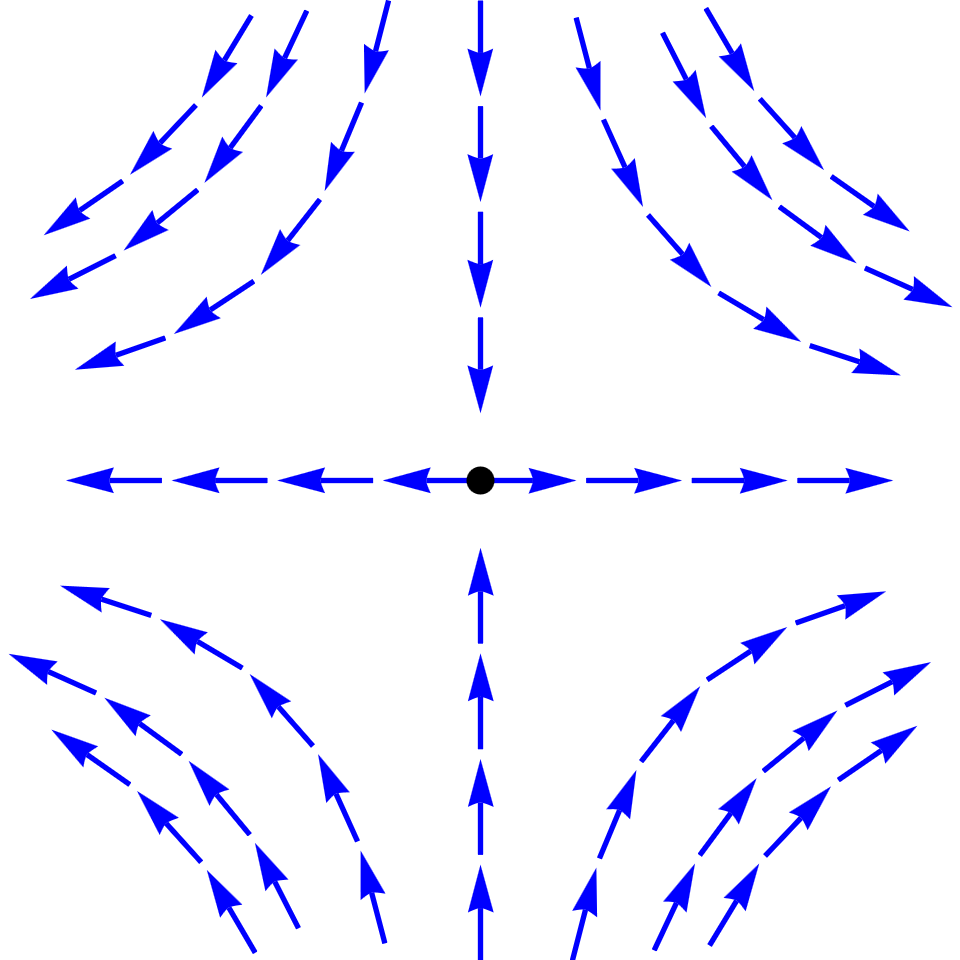}}
	\subfloat[$-1/2$ defect]
	{\includegraphics[width=0.28\textwidth]{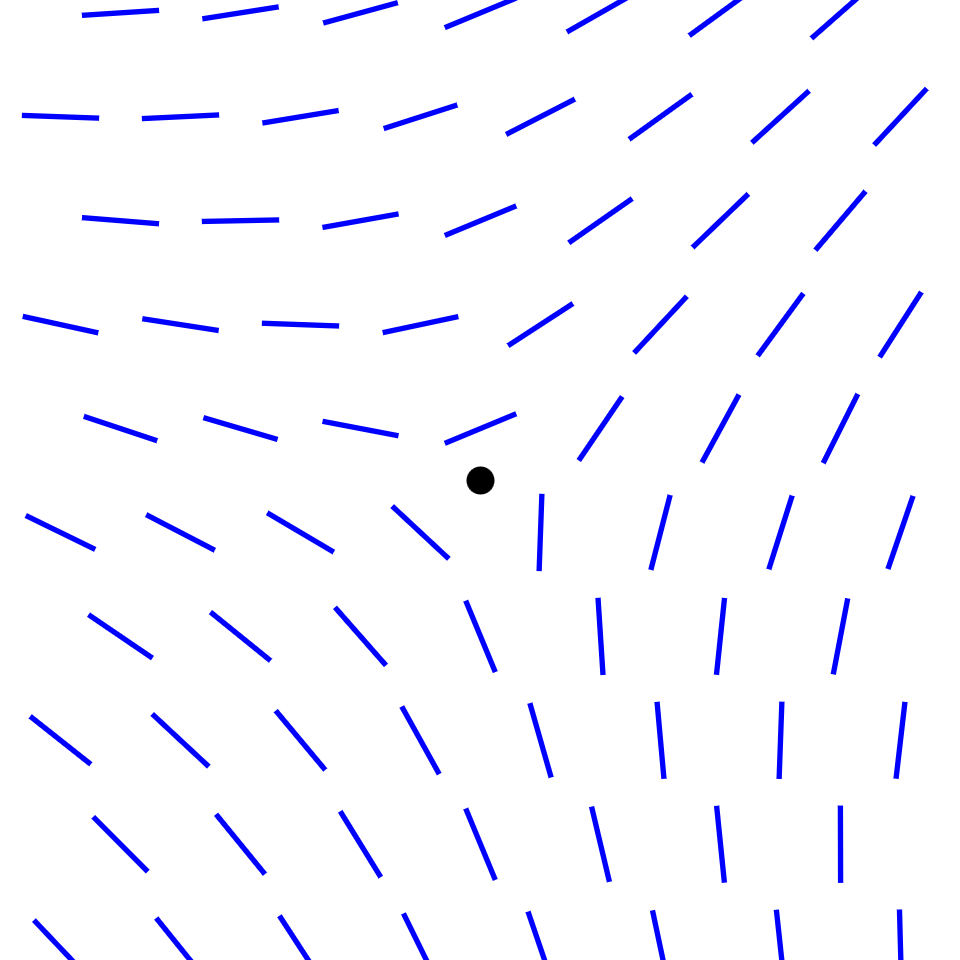}}
	\subfloat[$-1/3$ defect]
	{\includegraphics[width=0.28\textwidth]{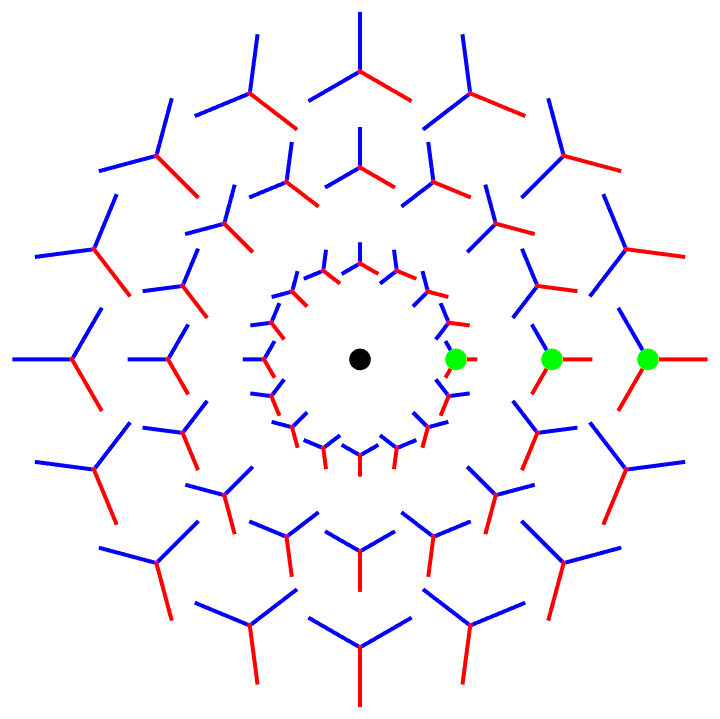}}
	\caption{Examples of topological defects. Top row: sketches of elementary positive defect textures for (a) $p=1$, (b) $p=2$, and (c) $p=3$. Bottom row: sketches of elementary negative defect textures for (d) $p=1$, (e) $p=2$, and (f) $p=3$. In (c) and (f), the red rod is colored to better visualize the rotation as it revolves once around the defect. Green dot is where the red rod returns to its starting position, after having rotated by $\pm 2\pi/3$, which is allowed because of the three-fold symmetry for a triatic ($p=3$) liquid crystal. Figure adapted from Ref.~\cite{vafa2023active}.}
	\label{fig:defectSketches}
\end{figure}

Topological defects are an important and fascinating feature of liquid crystals. In a liquid crystal, defects are regions where the orientational order of the molecules is disrupted, and are characterized by topological charges, which determine the degree and direction of rotation around the defect and the way in which the defects interact with one another. See Fig.~\ref{fig:defectSketches} for examples of topological defects. Defects evident when a bulk nematic is viewed through crossed polarizers are shown in the right panel of Fig.~\ref{fig:liquidCrystals}(a).

\begin{figure}[t]
	\centering
	\subfloat[]
	{\includegraphics[width=.4\columnwidth]{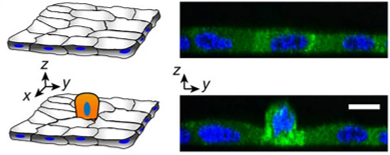}}
	\hfill
	\subfloat[]
	{\includegraphics[height=7cm]{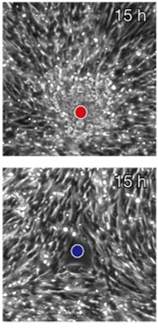}}
	\hfill
	\subfloat[]
	{\includegraphics[height=7cm]{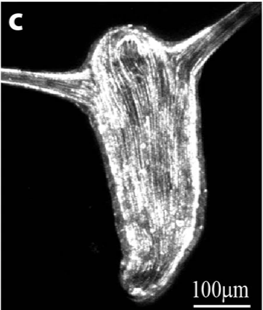}}
	\caption{Examples of functions for topological defects in biology. (a) Cell death (apoptosis) and extrusion can occur preferentially near defects in Madin-Darby canine kidney cells, adapted from Ref.~\cite{saw2017topological}. Cell mound formation in murine neural progenitor cells, adapted from Ref.~\cite{kawaguchi2017topological} (c) Limb formation in \emph{Hydra}, associated with regions of both positive and negative curvature, adapted from Ref.~\cite{maroudas2021topological}.}
	\label{fig:defectBiology}
\end{figure}

The study of disclinations has led to many fascinating discoveries in liquid crystal physics, including defects implicated in biological functions such as cell extrusion and apoptosis in mammalian epithelia~\cite{saw2017topological}, neural mound formation~\cite{kawaguchi2017topological}, bacterial competition~\cite{meacock2021bacteria}, and limb origination in the simple animal \emph{Hydra}~\cite{maroudas2021topological}, shown in Fig.~\ref{fig:defectBiology}.
For recent reviews on the significance of topological defects in biological systems, see e.g. Refs.~\cite{doostmohammadi2021physics, shankar2022topological, bowick2022symmetry}.

These biological examples where curved geometry plays a role motivates us to review here recent work on $p$-atic liquid crystals on curved surfaces~\cite{zhang2022fractional,vafa2022defectAbsorption}. Previous research studied ground state defect configurations on a sphere~\cite{lubensky1992orientational,nelson2002toward} and the interaction between liquid crystalline order and curved substrates~\cite{park1996topological,vitelli2004anomalous}. Turner and Vitelli~\cite{vitelli2004anomalous} further showed that inhomogeneous Gaussian curvature gives rise to an effective topological charge density, which, for a conventional (hyperbolic) cone, corresponds to negative (positive) topological charge concentrated at the apex. In this review, we focus on cones, the simplest example of curved geometry (flat everywhere except for a $\delta$-function curvature singularity at the apex).
 
This paper is organized as follows. We begin in Sec.~\ref{sec:order} by describing orientational order on disks and curved surfaces with various boundary conditions. After introducing a simple one-Frank-constant model in Sec.~\ref{sec:model}, we present ground state results for conventional cones in Sec.~\ref{sec:cone} and hyperbolic cones in Sec.~\ref{sec:hyperbolic}. The results for hyperbolic cone ground states are new, and go beyond the conventional cone results for both free and tangential boundary conditions presented in Refs.~\cite{zhang2022fractional} and \cite{vafa2022defectAbsorption}. We conclude in Sec.~\ref{sec:discussion} by summarizing our main results and commenting on extensions.

\section{Orientational order on disks and curved surfaces}
\label{sec:order}

Consider first an isolated $+1$ polar ($p=1$) defect on a flat disk with tangential boundary conditions that enforce a gradual rotation from $0$ to $2\pi$ at the boundary, as might occur in a Langmuir-Blodgett surfactant film with tilted boundary conditions~\cite{kaganer1999structure}. This situation is topologically stable. Now suppose we have an isolated $+1$ nematic ($p=2$) defect with similar boundary conditions. We would not expect this situation to be stable, since we can think of a $+1$ nematic defect being made of two $+1/2$ defects. Such defects, as discussed below, behave as point charges. We would then expect the two $+1/2$ defects to repel each other, once they are more than a core radius apart.

Now consider an isolated $+1$ polar defect on a curved surface, like a hemisphere with tangential boundary conditions: again, this situation is stable. Now suppose we have an isolated $+1$ nematic defect in a nematic film coating the surface. Since near a defect, the surface looks flat, then we would again expect an instability, since as before the two $+1/2$ defects that make up the $+1$ defect would repel each other.

\begin{figure}[t]
	\centering
	\includegraphics[width=\columnwidth]{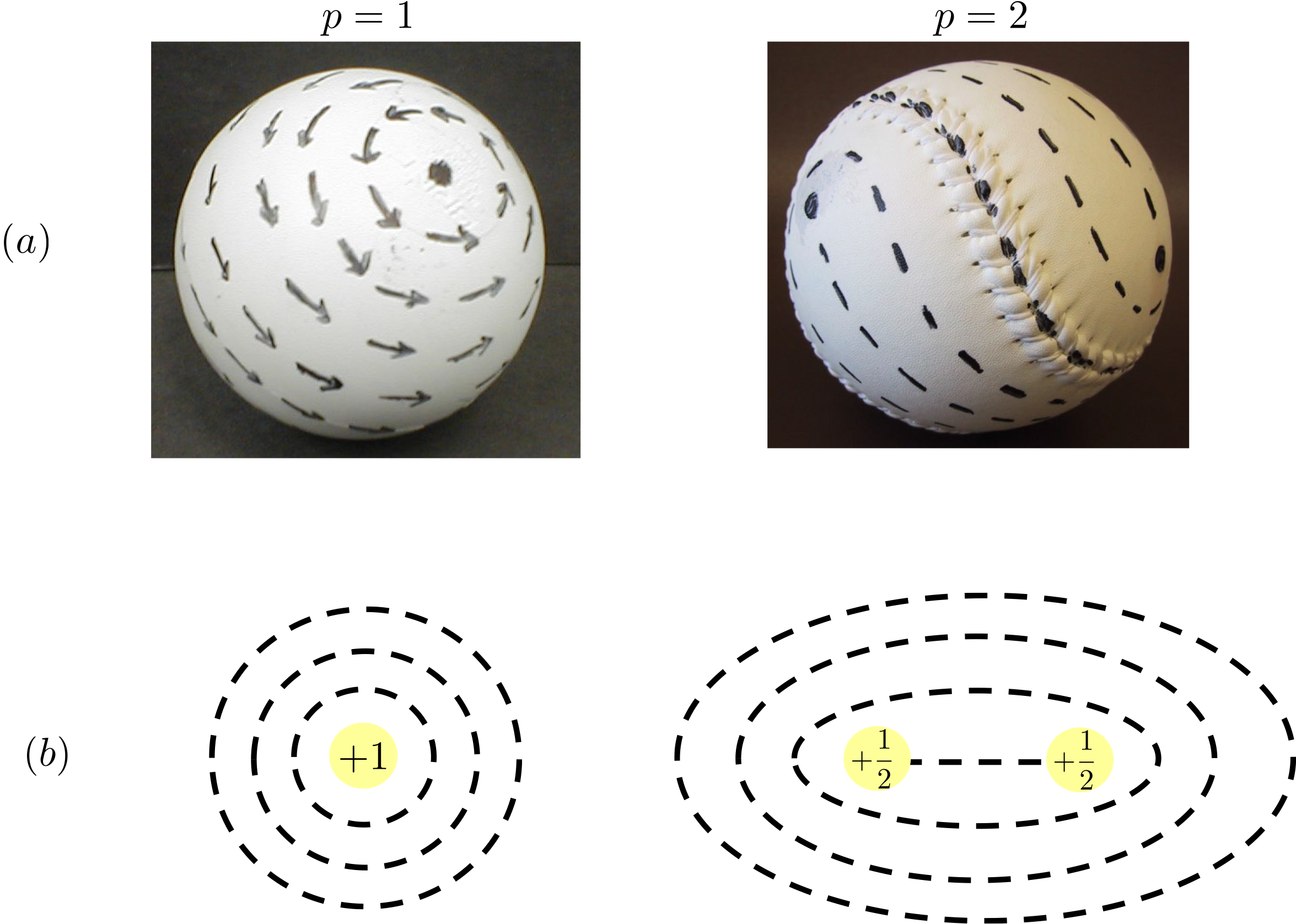}
	\caption{(a) Topological defects on spheres for polar ($p=1$) vs nematic ($p=2$) liquid crystals. (b) In nematic liquid crystals in a disk-like geometry constrained with boundary conditions to have a total topological defect charge of $+1$, a single $+1$ defect would prefer to split into two $+1/2$ defects, lowering the energy, due to the Coulombic repulsion.}
	\label{fig:comb}
\end{figure}

Lubensky and Prost~\cite{lubensky1992orientational} studied the equilibrium defect configurations on curved surfaces, including the sphere, which we illustrate in Fig.~\ref{fig:comb}. By the Gauss-Bonnet theorem~\cite{do1992riemannian}, the total charge of the defects on the sphere must be $2$. What does this mean? For polar order, let's say with two $+1$ defects, since defects repel each other, we would expect them to be at opposite poles. Now what about for nematic order? Here each $+1$ defect will split into a pair of $+1/2$ defects, leading to four $+1/2$ defects on the sphere sitting at the corners of an inscribed tetrahedron~\cite{lubensky1992orientational}. See Fig.~\ref{fig:comb} for a schematic.
 
Spheres, having constant Gaussian curvature everywhere, have been rather thoroughly analyzed. In the remainder of this work, we focus instead on cones and hyperbolic cones, which are flat everywhere except for at a single point (the apex), where there is a $\delta$-function of positive or negative Gaussian curvature singularity, respectively. One might imagine more exotic surfaces made of superpositions of cones and hyperbolic cones (defined more precisely below). In this case, cones and hyperbolic cones, and their associated texture response functions, can be considered as building blocks of complex geometries. Cones also appear beautifully in nature -- see Ref.~\cite{coral} for beautiful images of an Acropora monticulosacoral coral reef consisting of polyps patterned on conic surfaces.

\begin{figure}[t]
	\centering
	\subfloat[]
	{\includegraphics[width=\columnwidth]{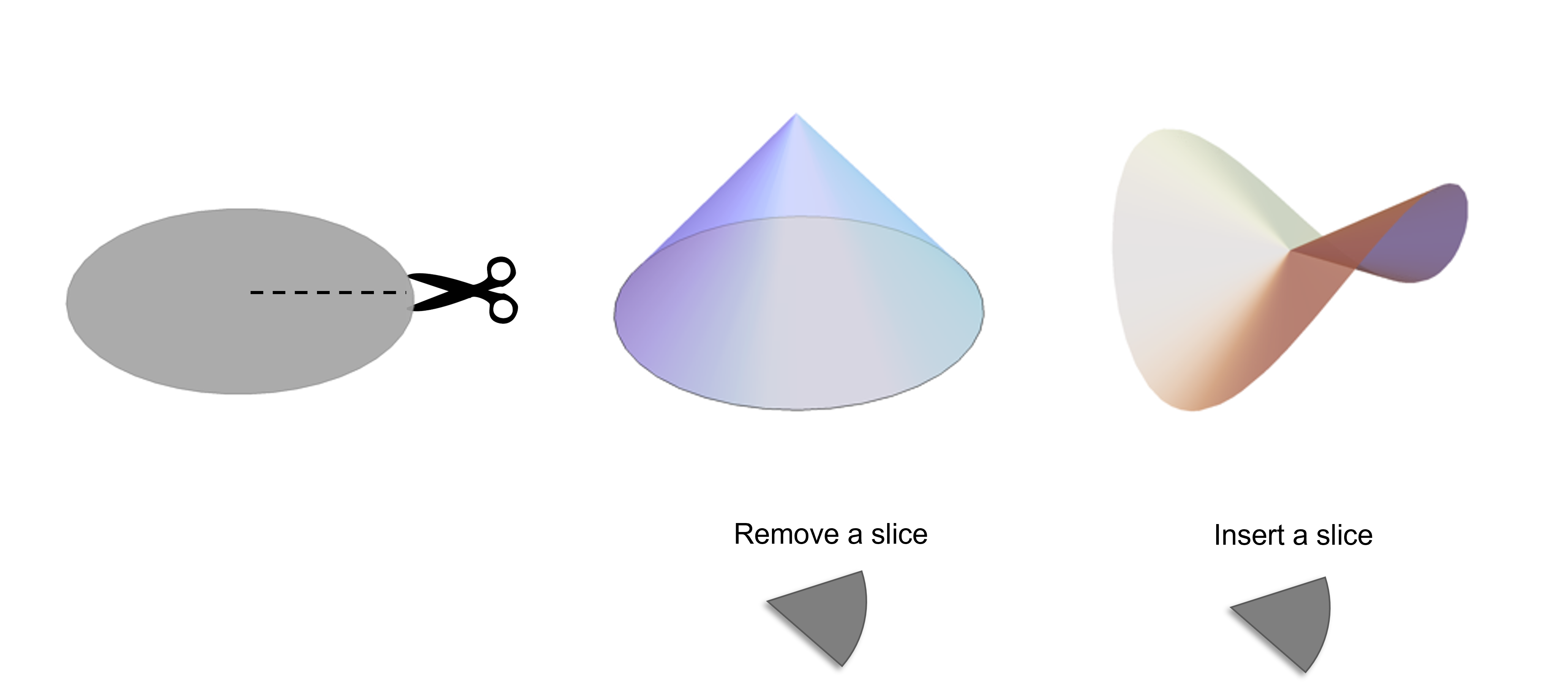}}
	\\
	\subfloat[]
	{\includegraphics[width=.95\columnwidth]{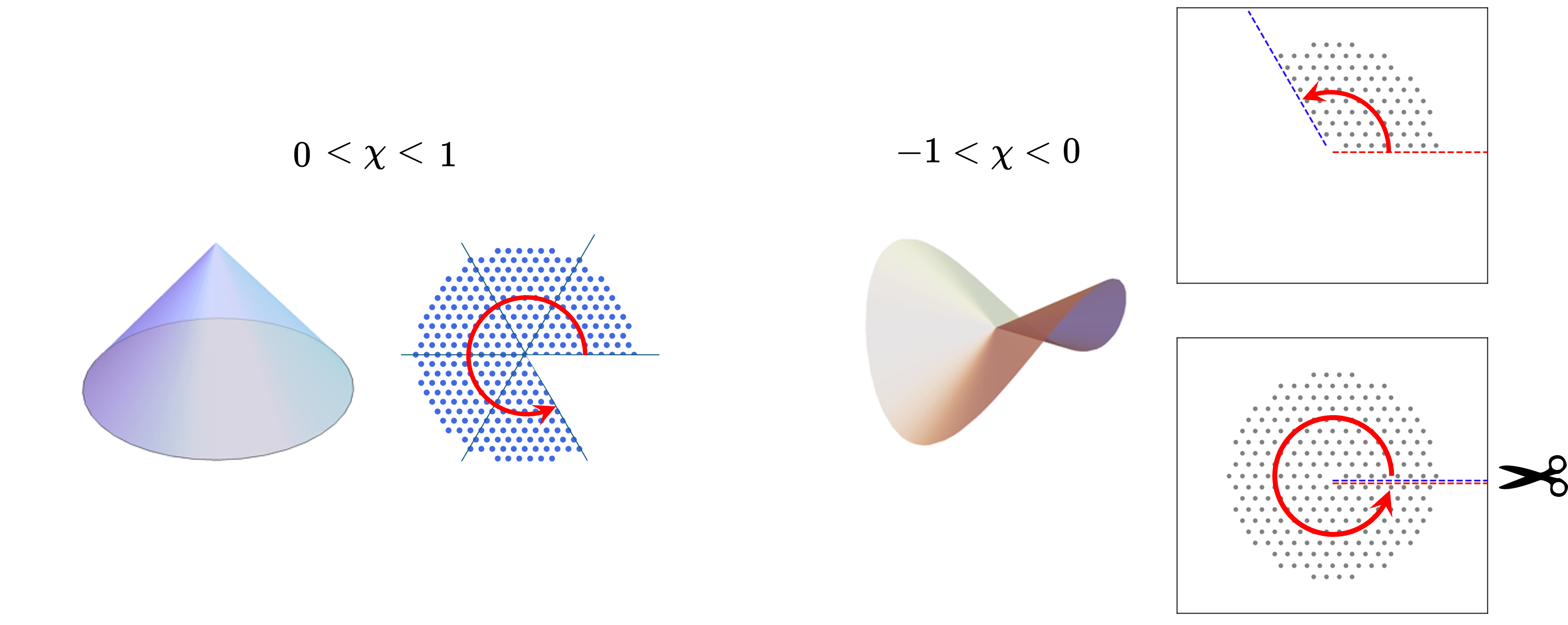}}
	\\
 
	\caption{(a) To construct a cone or hyperbolic cone from a disk, one can remove an angular slice from or insert an angular slice into the disk, respectively. (b) For simulations using the aforementioned construction, it is convenient to lay down a regular lattice of sites (in this case, a triangular lattice), on which individual liquid crystal molecules are placed, on a sector with angle $2 \pi (1 - \chi)$. For cones with $0 < \chi < 1$, the liquid crystal orientations on the two open boundaries of the angular sector after the slice removal are matched. Hyperbolic cones with $-1 < \chi < 0$ are composed of two angular sectors, one of which is a complete disk, with two sets of boundaries (marked by red and blue dashed lines) orientationally matched as shown on the right of the figure. As discussed in the text, using triangular or square lattices to do numerical energy minimizations on cones restricts our attention to special, discrete values of $\chi$.}
	\label{fig:cone-construction}
\end{figure}

What is the architecture of a cone? One can construct a cone by making a radial cut on a disk from its edge to the center, removing a wedge from a disk, and then smoothly gluing the two boundaries together. Similarly, one can construct a hyperbolic cone by inserting a wedge into the disk, and then gluing the four boundaries. In addition to the radius $R$ of the original disk, it is useful to characterize cones and hyperbolic cones by a dimensionless number $\chi$, the fraction of disk removed or inserted: $\chi = 0$ corresponds to a disk, $\chi > 0$ corresponds to a cone, and $\chi < 0$ corresponds to a hyperbolic cone. As discussed below, $-\chi$ can be interpreted as the effective apex topological charge~\cite{zhang2022fractional,vafa2022defectAbsorption}. See Fig.~\ref{fig:cone-construction} for how to construct cones and hyperbolic cones.

\begin{figure}[t]
	\centering
	\includegraphics[width = \columnwidth]{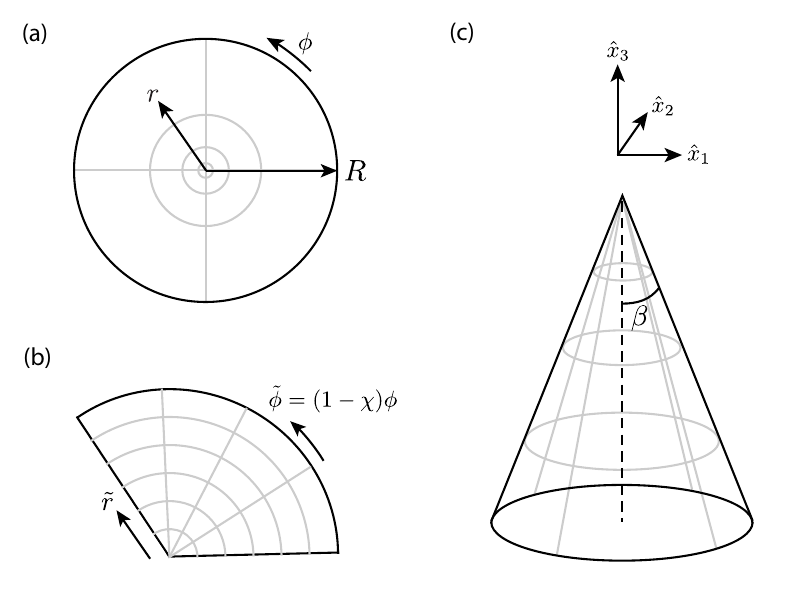}
	\caption{Schematic of the three coordinate systems for cones used in this paper: (a) Particularly useful and elegant are isothermal coordinates $z = re^{i\phi}$, which can be viewed as the result of squashing a cone into a plane, in a way that preserves the azimuthal angle $\phi$, where $0 \le \phi < 2\pi$, but distorts the radial coordinate $r$. Here $R$ is the maximum radius down the cone flanks in our isothermal coordinate system. (b) Also useful are the $\tilde z$ coordinates, the result of isometrically cutting open and unrolling a cone into a plane, so that the resulting azimuthal angle is now $\tilde\phi$, where $0 \le \tilde\phi < 2\pi(1-\chi)$. (c) Three-dimensional cone Cartesian coordinates $x_i$, where $\beta$ is the cone half-angle. Figure adapted from Ref.~\cite{vafa2022defectAbsorption}.}
	\label{fig:coordinates}
\end{figure}

A conventional cone can also described by the cone half-angle $\beta$, which is related to $\chi$ by $\chi = 1 - \sin\beta$. See Fig.~\ref{fig:coordinates} for a schematic of the three coordinate systems for cones used in this paper.

\section{Model}
\label{sec:model}

The $p$-atic texture, in a suitable choice of complex coordinates $z = r e^{i \phi}$ and $\bar z = r e^{-i \phi}$, can be described by a complex degree of freedom $Q = A e^{i\alpha}$, where $A$ is the amplitude and $\alpha$ is the phase. Deep in the ordered phase, where $|Q| \approx 1$, the Landau-de Gennes free energy $\mathcal F$ on a curved surface takes the form~\cite{vafa2022defectAbsorption}
\beq \mathcal F = J \int d^2z \left|\partial\alpha - i\left(\frac{p}{2}\right)\partial\varphi\right|^2, \label{eq:FSimple}\eeq
where $J$ is proportional to a liquid crystal Frank constant, and $\varphi(z, \bar z)$ is a fixed function that describes the geometry. In particular, $\varphi$ satisfies
\beq \nabla^2 \varphi = -2 G ,\eeq
where $G$ is the Gaussian curvature, which for us is proportional to a $\delta$-function. In the special case of a conventional cone or hyperbolic cone, $\varphi = -\chi \ln (z \bar z)$, where $0 < \chi < 1$ describes cones and $-1 < \chi < 0$ describes their hyperbolic counterparts.

\begin{figure}[t]
	\centering
    \subfloat[]
	{\includegraphics[width=.6\columnwidth]{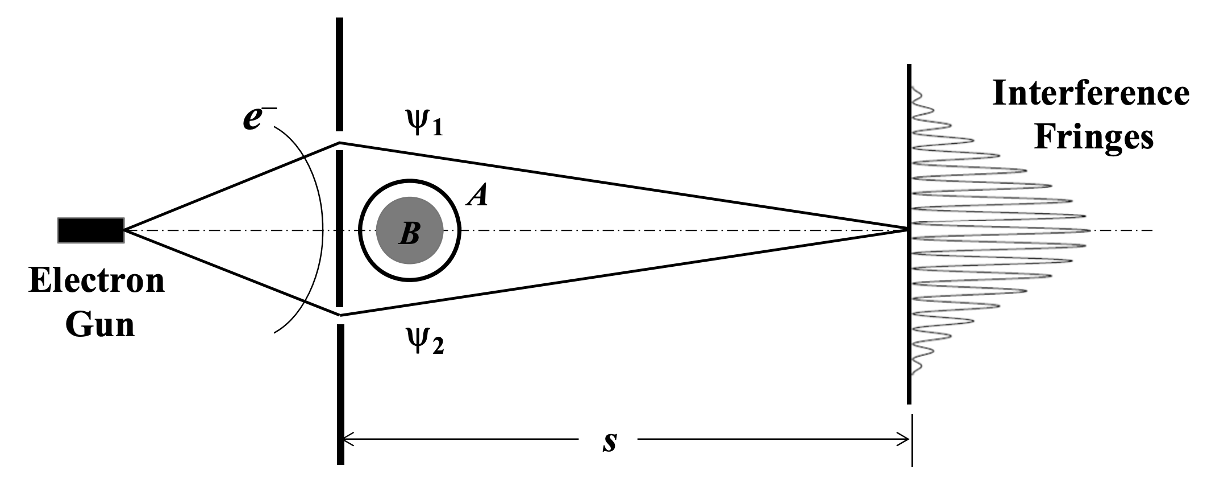}}
    \subfloat[]
    {\includegraphics[width=.4\columnwidth]{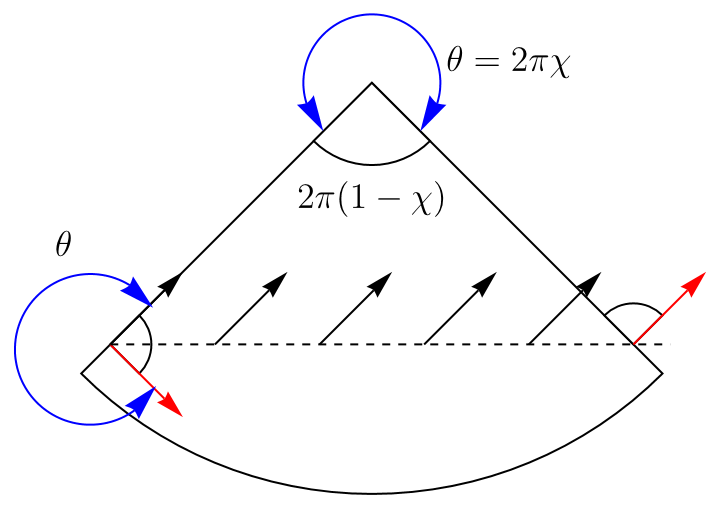}}
	\caption{(a) Illustration of the Aharonov-Bohm effect, where, depending on the path of the electron, the phases $\Psi_1$ and $\Psi_2$ are related by $\Psi_1 = e^{iq\Phi} \Psi_2$ and the magnetic flux $\Phi = \int \vec B \cdot \vec n da$, where $da$ is an area element. Figure adapted from Ref.~\cite{kasunic2019magnetic}. (b) An example of the parallel transport of a vector on a conical geometry in the unrolled coordinates, where the left and right edges of the wedge are identified. As the black vector is parallel transported from the left edge to the right edge on this locally flat surface, it effectively rotates by $\theta = 2\pi \chi$ so there is a mismatch when the cone is reassembled. On a general curved geometry, a vector, after traversing a closed loop $C$, rotates by $\theta = \int G da$, where the region of integration is bounded by $C$ and $G$ is the Gaussian curvature.  In connection to the Aharonov-Bohm effect, the curvature acts as the magnetic field, and explicitly, $\theta$ for the cone is the analog of $q\Phi$ for the Aharonov-Bohm effect, where $q$ is the electron charge. The case of a cone with a sharp tip corresponds to an infinitesimally thin solenoid. A decapitated cone resembles a solenoid of finite radius.}
	\label{fig:Aharonov-Bohm}
\end{figure}

Eq.~\eqref{eq:FSimple} resembles a London model of a superfluid with phase angle $\alpha(z)$ on a flat surface, with a vector potential $\frac{p}{2} \partial \varphi$. If we compare this to the Aharonov-Bohm effect, the curvature acts as the analogue of the magnetic field; in particular, the case of a cone corresponds to an infinitesimally thin solenoid. See Fig.~\ref{fig:Aharonov-Bohm} for a reminder of the Aharonov-Bohm effect, and its connection with cone geometries.

In the following two sections, we present our results for first conventional cones and then hyperbolic cones.

\section{Conventional Cones}
\label{sec:cone}

\begin{figure}[t]
	\centering
	\includegraphics[width=0.5\columnwidth]{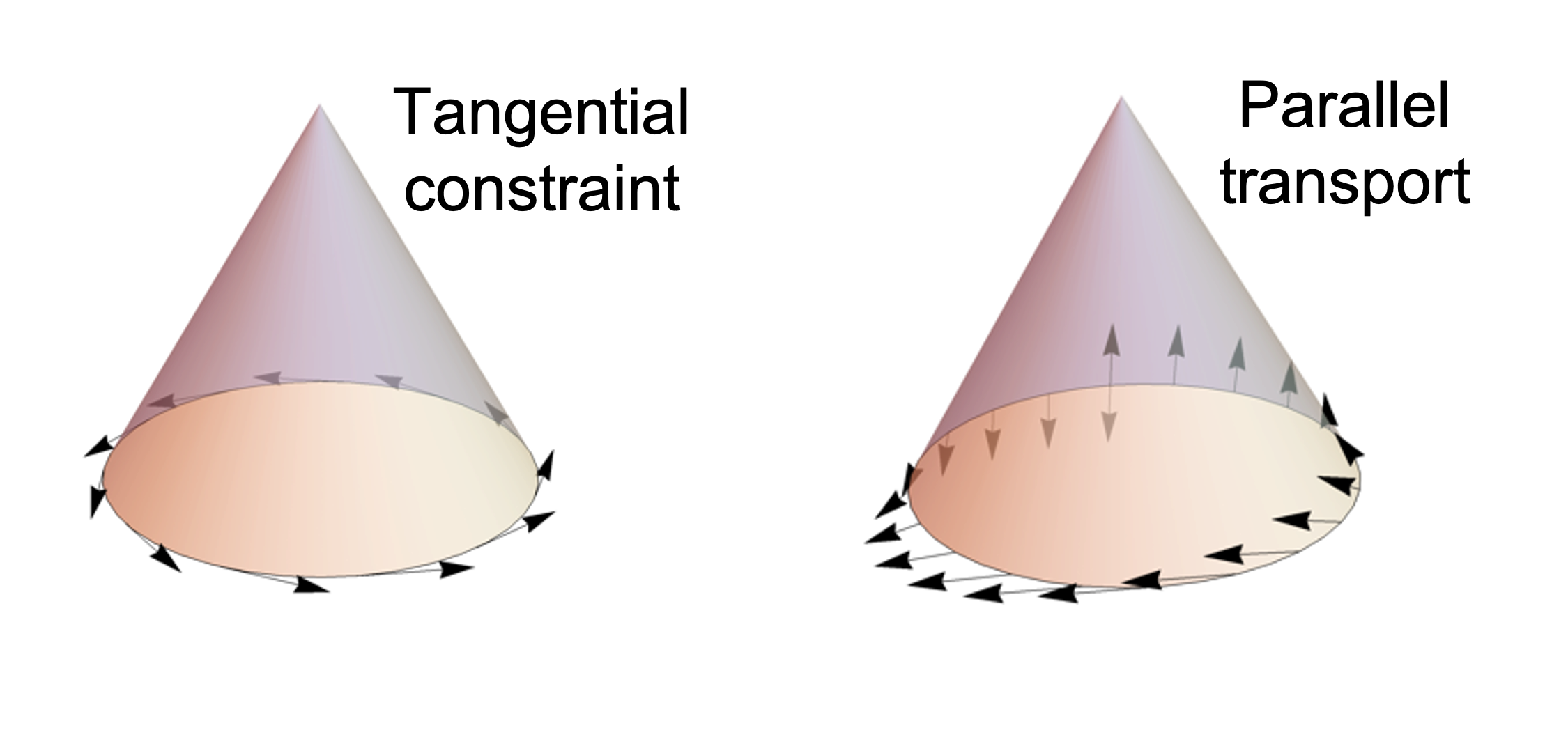}
	\caption{Orientation of the liquid crystal direction vectors as constrained by tangential boundary conditions (left) versus as aligned via parallel transport (right) on a $\chi = 1/2$ cone, corresponding to a cone half-angle $\beta = 30^\circ$ (i.e., $1-\sin\beta = 1/2$). Parallel transport of the orientation vector is typically (as is the case for $p=1$ in this figure) incompatible with free boundary conditions, because the order parameter is discontinuous after going around the circumference. Parallel transport is always incompatible with tangential boundary conditions. This incompatibility imposes geometric frustration on the $p$-atic texture.}
	\label{fig:tangential_BC-frustration}
\end{figure}

We first describe the role of boundary conditions. In this paper, we consider two boundary conditions at the base of the cone: free and tangential. In free boundary conditions, the director is free, and the number of topological defects is not fixed, whereas in tangential boundary conditions, the director is tangential to the boundary. The result of this constraint is that for tangential boundary conditions, the total charge of defects must be $+1$, whereas there is no such constraint for free boundary conditions. Tangential boundary conditions are more complicated than free boundary conditions since the tangential boundary conditions are incompatible with parallel transport, as illustrated in Fig.~\ref{fig:tangential_BC-frustration}.

Free boundary conditions are straightforward to implement as there is no such constraint on the total charge: defects can easily enter and leave at the cone base.  We first present the results for free boundary conditions~\cite{zhang2022fractional}, and then for the more subtle problem of $p$-atic liquid crystals with tangential boundary conditions~\cite{vafa2022defectAbsorption}.

\subsection{Free boundary conditions}

For $p$-atics on cones with free boundary conditions, studied in Ref.~\cite{zhang2022fractional}, the ground state configurations have integer $s_0$ defects of charge $1/p$ at the apex to minimize the magnitude of the total effective apex charge 
\beq \label{eq:qA_free}
q_A = -\chi + s_0/p,
\eeq 
where $s_0$ is  chosen to minimize the apex charge. Thus, $s_0$ is given by~\cite{zhang2022fractional}
\beq \label{eq:ground_free}
s_0 = \underset{s}{\mathrm{argmin}} |-\chi + s/p| .
\eeq
Other values of $s$ describe metastable states~\cite{zhang2022fractional}.
\begin{figure}[t]
	\centering
	{\includegraphics[width=\columnwidth]{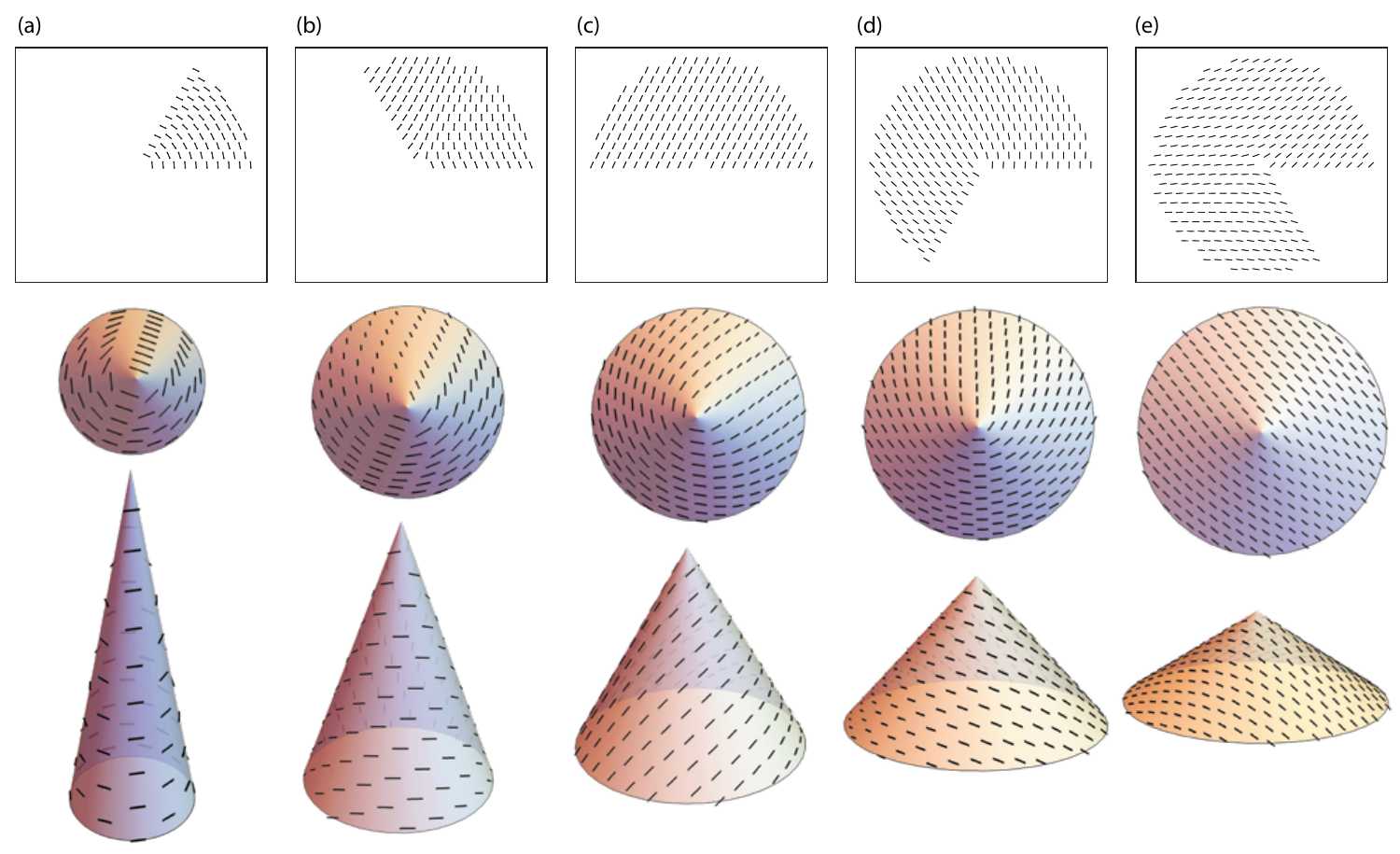}}
	\caption{Ground state configurations of $p=2$ nematic liquid crystals placed on the surfaces of cones commensurate with a perfect triangular lattice mesh, with cone angles such that $\sin\beta = 1/6$ (a), $2/6$ (b), $3/6$ (c), $4/6$ (d), $5/6$ (e), corresponding to $\chi = 5/6$, $4/6$, $3/6$, $2/6$, and $1/6$, respectively. The top row shows the configurations isometrically projected onto the polar plane, while the middle and bottom rows show the texture on the surface of the cones in three dimensions. Figure adapted from Ref.~\cite{zhang2022fractional}.}
	\label{fig:free_BC_results}
\end{figure}

Fig.~\ref{fig:free_BC_results} shows the ground state textures obtained by numerical energy minimizations for nematic ($p=2$) liquid crystals with free boundary conditions for cone angles commensurate with a perfect triangular lattice. The resulting ground state apex charge $q_A$ on cones with angles commensurate with either triangular or square lattices are summarized in Table~\ref{tab:free_cone}.

\begin{table}[htb]
    \begin{tabular}{c|c|c|c|c|c|c}
$\chi$ & $p=1$ & $p=2$ & $p=3$ & $p=4$ & $p=5$ & $p=6$ \\
\hline$\frac{1}{6}$ & $-\frac{1}{6}$ & $-\frac{1}{6}$ & $\pm \frac{1}{6}$ & $\frac{1}{12}$ & $\frac{1}{30}$ & 0 \\
$\frac{1}{4}$ & $-\frac{1}{4}$ & $\pm \frac{1}{4}$ & $\frac{1}{12}$ & 0 & $-\frac{1}{20}$ & $\pm \frac{1}{12}$ \\
$\frac{2}{6}$ & $-\frac{1}{3}$ & $\frac{1}{6}$ & 0 & $-\frac{1}{12}$ & $\frac{1}{15}$ & 0 \\
$\frac{3}{6}\left(\frac{2}{4}\right)$ & $\pm \frac{1}{2}$ & 0 & $\pm \frac{1}{6}$ & 0 & $\pm \frac{1}{10}$ & 0 \\
$\frac{4}{6}$ & $\frac{1}{3}$ & $-\frac{1}{6}$ & 0 & $\frac{1}{12}$ & $-\frac{1}{15}$ & 0 \\
$\frac{3}{4}$ & $\frac{1}{4}$ & $\pm \frac{1}{4}$ & $-\frac{1}{12}$ & 0 & $\frac{1}{20}$ & $\pm \frac{1}{12}$ \\
$\frac{5}{6}$ & $\frac{1}{6}$ & $\frac{1}{6}$ & $\pm \frac{1}{6}$ & $-\frac{1}{12}$ & $-\frac{1}{30}$ & 0
\end{tabular}
\caption{Ground state apex defect charge $q_A$ according to Eqs.~\eqref{eq:qA_free} and \eqref{eq:ground_free} for cone angles commensurate with either triangular or
square lattices, in agreement with the numerics. Square lattice meshes on cone flanks allow efficient evaluations of the ground stat energy for cones with $\chi = 1/4$, $1/2$, and $3/4$. Triangular lattice meshes allow similar efficient evaluations for cones with $\chi = 1/6$, $2/6$, $3/6$, $4/6$, and $5/6$.}
\label{tab:free_cone}
\end{table}

\begin{figure}[t]
	\centering
	{\includegraphics[width=.5\columnwidth]{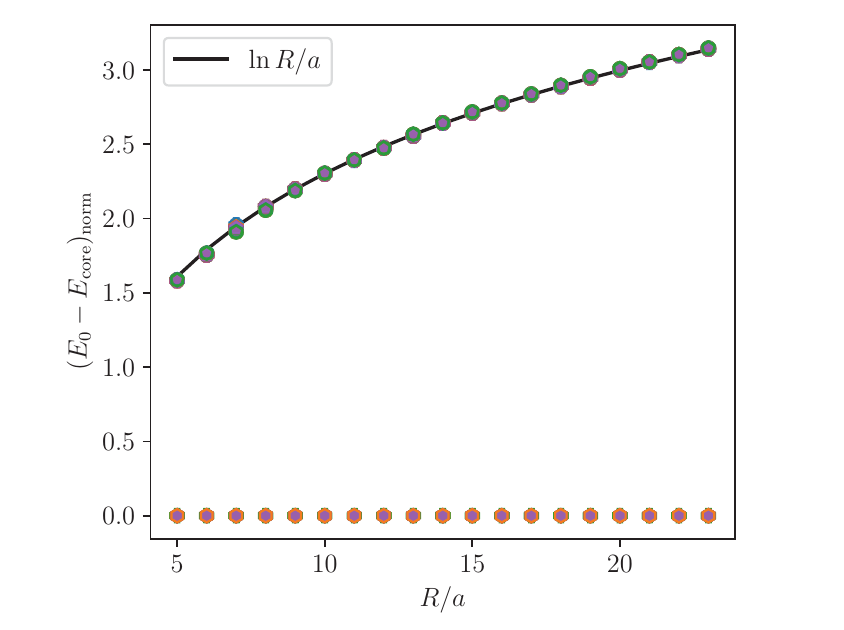}}
	\caption{The ground state energy of $p$-atics on conventional cones with free boundary conditions as a function of system size. Colored dots are ground state energies (with a core energy subtracted off) extracted from numerical energy minimizations of cones with different commensurate angles on square and triangular lattices and scaled according to Eq.~\eqref{eq:energy_scaled}. Black line is $\ln(R/a)$. The energies for entries with $q_A = 0$ in Table~\ref{tab:free_cone} (red dots) are independent of $R$ in the ground state. Figure adapted from Ref.~\cite{zhang2022fractional}.}
	\label{fig:collapse}
\end{figure}

The ground state energies for different cone angles and values of $p$ typically scale logarithmically as a function of system size and collapse onto the same curve when scaled appropriately (see Fig.~\ref{fig:collapse}),
\beq 
(E_0-E_{\text {core }})_\mathrm{norm} \equiv \frac{E_0-E_{\text {core }}}{2 \pi \tilde{J} p^2 q_A^2 / (1 - \chi)}=\ln (R / a) \label{eq:energy_scaled},
\eeq 
where $E_0$ is the energy of the ground state configuration, $E_\mathrm{core}$ is a constant core energy that depends on both $p$ and $q_A$ (and microscopic details like the lattice structure), $a$ is the lattice constant, $R$ is the total radial length of the cone, and $\tilde J$ is a coupling constant aligning the $p$-atic order parameter that depends on the lattice geometry~\cite{zhang2022fractional}. Exceptions to the logarithmically diverging ground state energy arise when $q_A(p, \chi) = 0$, as happens for 11 out of 42 of the entries in Table~\ref{tab:free_cone}. See Fig.~\ref{fig:collapse}.

\subsection{Tangential boundary conditions}

To enforce the tangential boundary conditions, just as in electrostatics, for each defect charge inside the rim at position $z$, we place a like-signed image charge outside the rim at position $\tilde z = R^2/\bar z$, see Fig.~\ref{fig:image-charges}. (See the Appendix of Ref.~\cite{vafa2022defectAbsorption} for worked examples; to impose free boundary conditions with defects on the cone flanks, opposite-signed image charges must be used.)

\begin{figure}[t]
	\centering
	\includegraphics[width=\columnwidth]{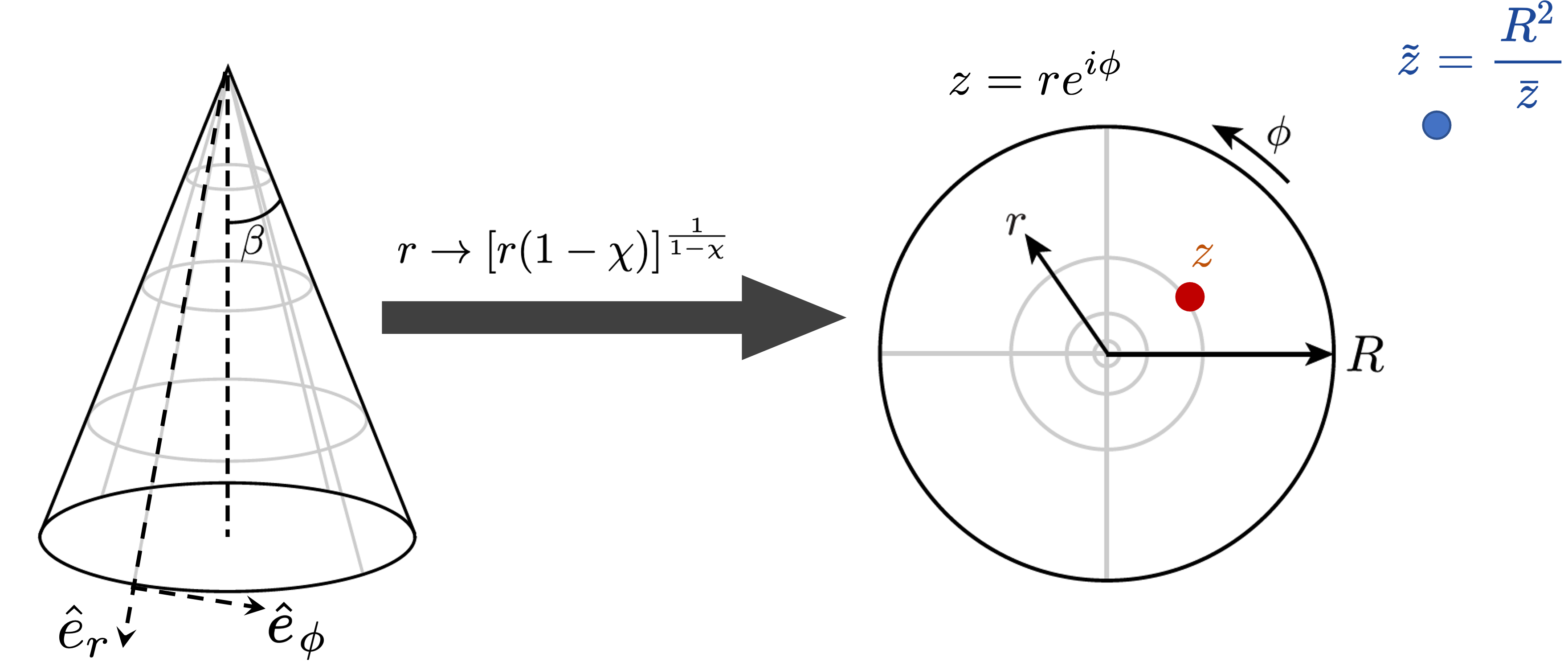}
	\caption{To enforce tangential boundary conditions, the conic surface is first transformed to isothermal coordinates, and then a like-signed image charge is placed outside the cone rim for every defect charge inside the cone rim.}
	\label{fig:image-charges}
\end{figure}

The goal is to find the configuration of defect charges that minimizes the free energy. Since the apex charge $-\chi$ is negative for conventional cones, which, depending on the deficit angle, allows some of the defects to be eaten by the geometrical charge at the apex. The remaining defects sit equidistant along a ring on the flank~\cite{vafa2022defectAbsorption}. There are three types of defect configurations:
\begin{enumerate}
	\item For sharp cones (large deficit angle), all of the $+1/p$ defects can be eaten by the geometrical charge at the apex.
	\item For shallow cones (intermediate deficit angle), only some of the $+1/p$ defects are at the cone apex; the rest remain equally spaced around a concentric ring on the cone flank, at a distance that minimizes the total energy.
	\item In the case of a disk (zero deficit angle), all of the $+1/p$ defects are equally spaced along a concentric ring inside the disk.
\end{enumerate}
To find the number of such defects and their positions, we compute the free energy (incorporating the image charges to enforce the tangential boundary conditions), giving~\cite{vafa2022defectAbsorption}

\beq
\mathcal F = -\pi\frac{p^2}{2} J\biggl\{\sum_{m<n}\sigma_m\sigma_n\biggl[
\underbrace{\ln \frac{|z_m - z_n|^2}{R^2}}_{\color{Red}\mathclap{\substack{\text{\normalsize Repulsion from} \\ \text{\normalsize other defect}}}}
+
\underbrace{\ln \left|1 - \frac{z_m\overline{z_n}}{R^2}\right|^2}_{\color{Green}\mathclap{\substack{\text{\normalsize Repulsion from} \\ \text{\normalsize other defect's} \\ \text{\normalsize image}}}}
\biggr]
+ \sum_j
\underbrace{\sigma_j^2\ln\left(1 - \frac{|z_j|^2}{R^2}\right)}_{\color{Goldenrod}\mathclap{\substack{\text{\normalsize Repulsion from} \\ \text{\normalsize one's own image}}}}
- 
\chi \sum_j \biggl(
\underbrace{\sigma_j}_{\vstrut{2ex}\color{Plum}\mathllap{\substack{\text{\normalsize Geometric} \\ \text{\normalsize attraction} \\ \text{\normalsize to apex}}}}
\underbrace{\frac{\sigma_j^2}{2}}_{\color{SkyBlue}\mathrlap{\substack{\text{\normalsize Self energy} \\ \text{\normalsize on a non-flat} \\ \text{\normalsize surface}}}}
\biggr)\ln \frac{|z_j|^2}{R^2} \biggr\}\label{eq:FCone}. \eeq

\begin{figure}[t]
	\centering
	\includegraphics[width=\columnwidth]{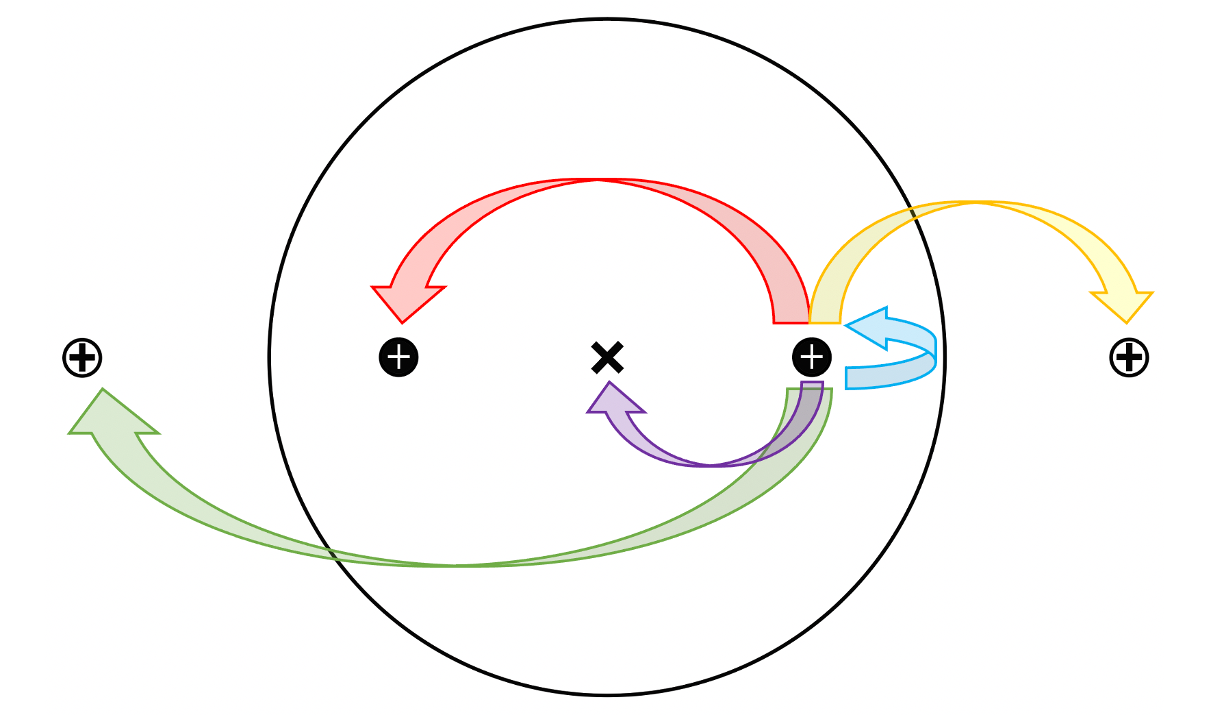}
	\caption{Schematic representation of the various contributions to the energy in Eq.~\eqref{eq:FCone} for a $p=2$ liquid crystal on a cone with tangential boundary conditions, with one geometrical charge at the apex (marked by the black $\times$), two $+1/2$ defects on the flanks, and two image charges outside the rim. The colors of the arrows representing the pairwise interactions correspond to the colored labels of the terms in Eq.~\eqref{eq:FCone}.}
	\label{fig:graphical-energy}
\end{figure}

See Fig.~\ref{fig:graphical-energy} for a graphical interpretation of the above equation.

We now assume in particular that $k$ defects of charge $+1/p$ are equally spaced on a ring at a distance $d=xR$ on the cone flank, i.e., these defects have complex coordinates $z_j = d e^{2\pi i (j/k)}$, $j=0,\ldots,k-1$, and the remaining $p-k$ defects are absorbed by the cone apex. Then the free energy becomes (up to a constant)
\beq \label{eq:FConeMin}
\mathcal F = -\pi\frac{p^2}{2} J \left[\frac{1}{p^2}\frac{k(k-1)}{2}\ln x^2 + \frac{k}{p^2}\ln(1 - x^{2k}) + k\frac{\chi'}{p}\ln x^2\right],
\eeq
where 
\beq 
\chi' = - \left(1 - \frac{1}{2p}\right)\chi + \frac{p-k}{p} .
\eeq
The $\chi'$ term determines whether a defect is absorbed by the core. These transitions happen at a set of critical cone angles such that
\beq 
\chi_c'=0 \implies \chi_c = \frac{2(p-k)}{2p-1} ,\label{eq:chi_c}
\eeq
and $\mathcal F$ turns out to be minimized when 
\beq 
x = d/R = \left(\frac{k - 1 +2p\chi'}{3 k - 1 +2p \chi'}\right)^{\frac{1}{2k}} , \label{eq:dxR}
\eeq
where $k$ is chosen such that
\beq  
\chi' - 1/p < 0 \le \chi' ,
\eeq
or equivalently in terms of the cone apex charge $-\chi$,
\beq 
\frac{2(p-k-1)}{2p-1} \le \chi < \frac{2(p-k)}{2p-1} .
\eeq

\begin{figure}[t]
	\centering
	\includegraphics[width=0.9\columnwidth]{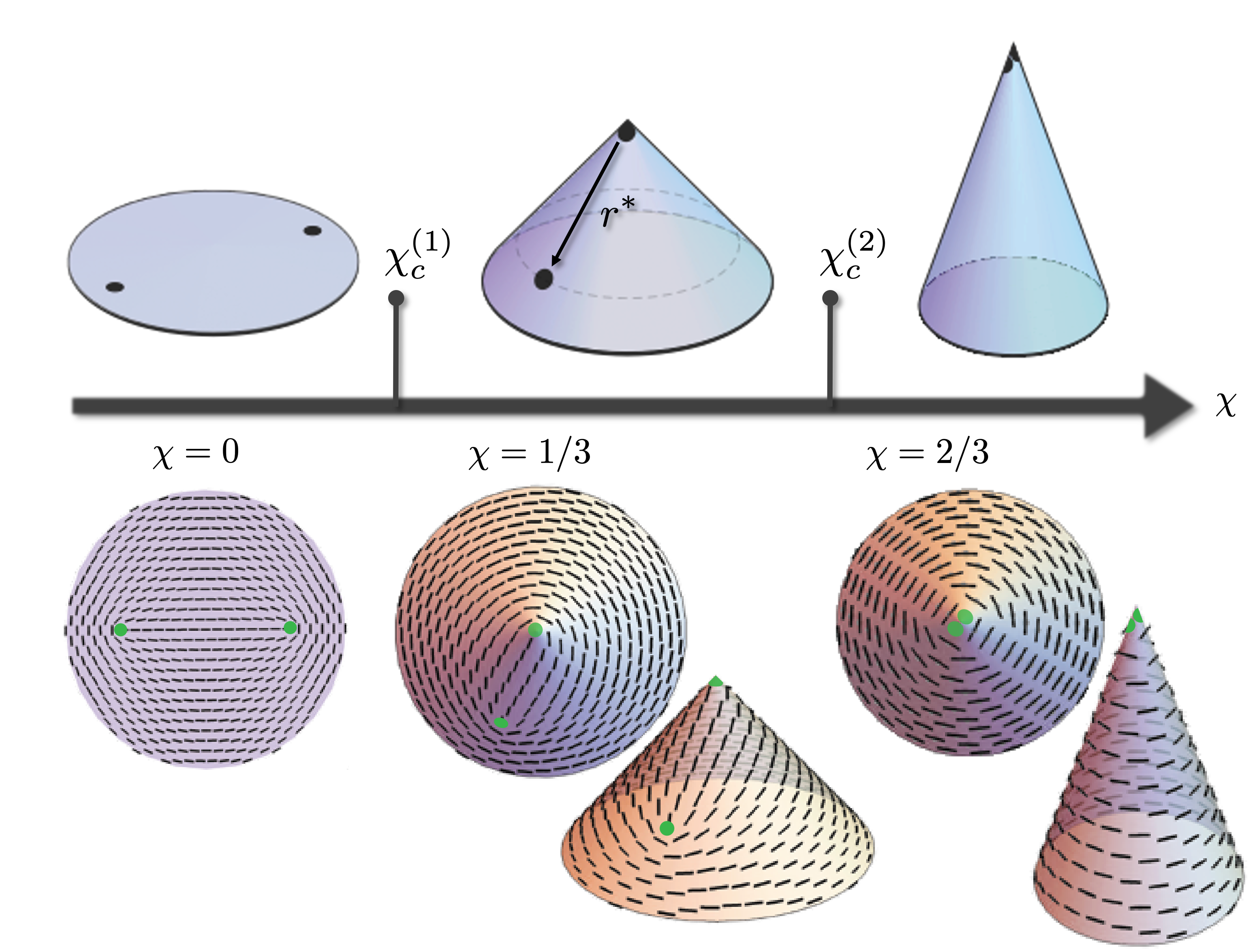}
	\caption{Illustration of defect absorption transitions as a function of increasing $\chi=1-\sin\beta$, where $\beta$ is the cone half-angle. For $p=2$ liquid crystals under tangential boundary conditions, there are two such transitions, as given by Eq.~\eqref{eq:chi_c} when $k = 1, 2$. As can be seen from Fig.~\ref{fig:cone-agreement}, $\chi_c^{(1)} = 0$, so one flank defect is absorbed as soon as $\beta$ drops below its disk value of $90^\circ$. Ground state numerical textures are shown on the bottom for three different values of $\chi$. On a flat disk ($\chi = 0$), there are two $+1/2$ defects, labeled with green dots, at positions given by Eq.~\eqref{eq:dxR} with $p=2$. On the surface of a cone corresponding to $\chi = 1/3 > \chi_c^{(1)} = 0$, there is one $+1/2$ defect on the flank and another at the cone apex. On a cone corresponding to $\chi \ge 2/3$, there are always two $+1/2$ defects at the cone apex, leaving none on the flanks. From Fig.~\ref{fig:cone-agreement}, we see that $\chi_c^{(2)} = 2/3$.
	}
	\label{fig:nematic_textures}
\end{figure}

\indent See Fig.~\ref{fig:nematic_textures} for a graphical illustration of defect absorption transitions for $p=2$ as well as the corresponding nematic liquid crystal textures.

\begin{figure}[t]
	\centering
	\subfloat[]
	{\includegraphics[width=1.0\columnwidth]{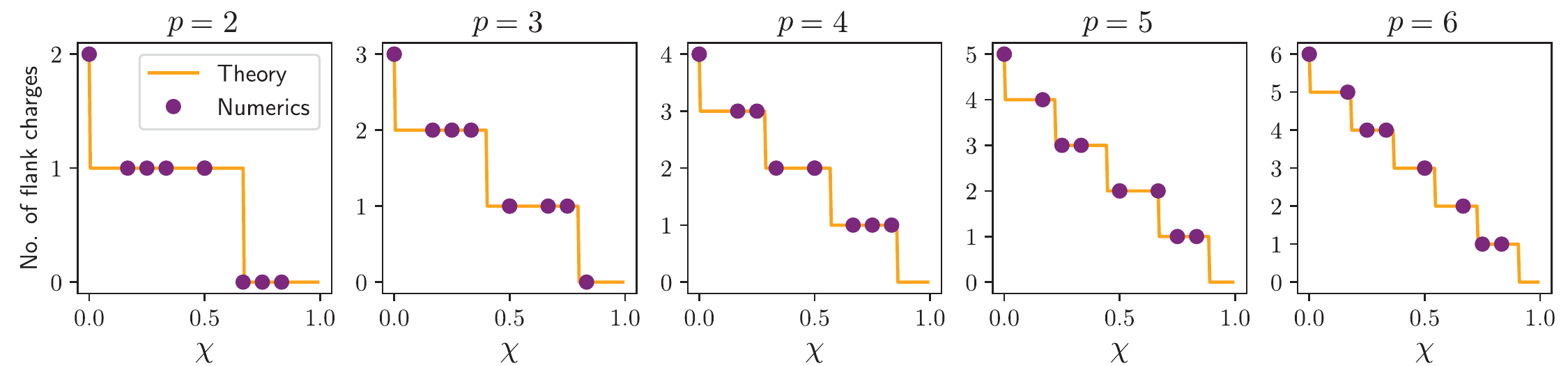}}
	\\
	\subfloat[]
	{\includegraphics[width=1.0\columnwidth]{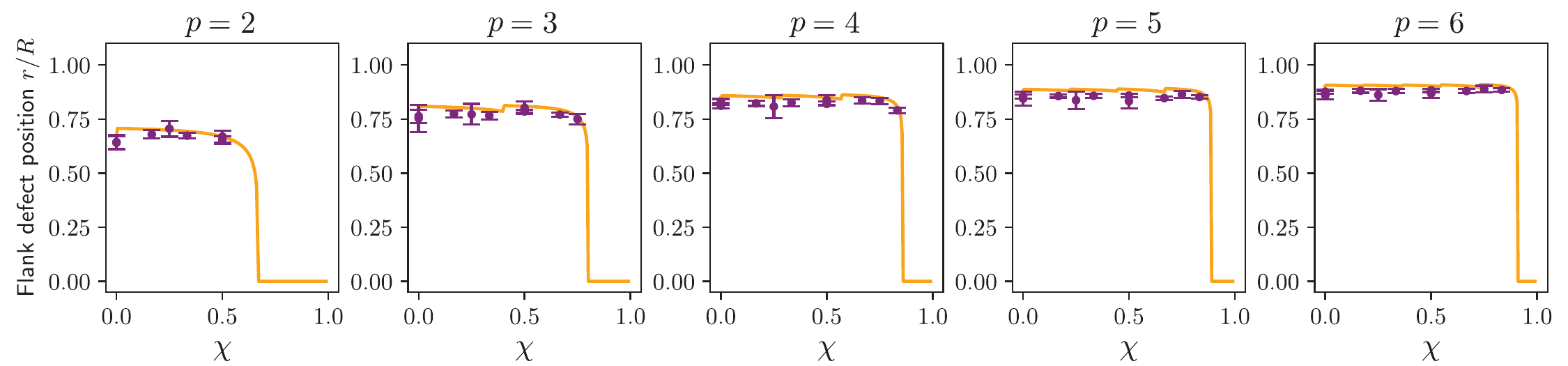}}
	\caption{Top row: plots of number of flank charges as a function of $\chi=1-\sin\beta$, where $\beta$ is the cone half-angle. Purple dots are from numerical energy minimization, and orange line is theoretical prediction for defect absorption transitions (Eq.~\eqref{eq:chi_c}). Bottom row: plots of flank defect positions in the ground states of $p$-atics on cones as a function of $\chi$. Purple markers are from numerical energy minimization, and orange curve is theoretical prediction (Eq.~\eqref{eq:dxR}). The theory predicts small jumps in the defect positions where, for example, the defect number changes from $2$ to $1$ when $p=3$. Our numerical minimizations cannot resolve these tiny jumps.}
	\label{fig:cone-agreement}
\end{figure}

Finally, see Fig.~\ref{fig:cone-agreement} for comparison of analytical results to simulations for a discretized lattice model of a $p$-atic liquid crystal - the agreement is excellent.

\section{Hyperbolic cones}
\label{sec:hyperbolic}

Having reviewed conventional cones~\cite{zhang2022fractional,vafa2022defectAbsorption}, we now present new results for hyperbolic cones, which allow access to negative values of $\chi$, as shown in Fig.~\ref{fig:cone-construction}. We first consider the case of free boundary conditions, and then tangential boundary conditions.

\subsection{Free boundary conditions}

For free boundary conditions, we expect that $|q_A|$ will still be minimized (see Eq.~\eqref{eq:qA_free}). Since now $\chi < 0$, this corresponds to the number of absorbed or emitted $s_0$ defects changing sign compared to the case for a conventional cone. See Tables~\ref{tab:free_cone} and \ref{tab:hyperbolic_cone} for a comparison of the ground state apex charges of hyperbolic cones to those of conventional cones. Note that the apex charges simply change sign as one moves from positive to negative values of $\chi$. The symmetry reflects the fact that the physics is invariant to an overall sign flip of all the charges (geometrical and topological). 

\begin{table}[htb]
    \begin{tabular}{c|c|c|c|c|c|c}
$\chi$ & $p=1$ & $p=2$ & $p=3$ & $p=4$ & $p=5$ & $p=6$ \\
\hline$-\frac{1}{6}$ & $\frac{1}{6}$ & $\frac{1}{6}$ & $\pm \frac{1}{6}$ & $-\frac{1}{12}$ & $-\frac{1}{30}$ & 0 \\
$-\frac{1}{4}$ & $\frac{1}{4}$ & $\pm \frac{1}{4}$ & $-\frac{1}{12}$ & 0 & $\frac{1}{20}$ & $\pm \frac{1}{12}$ \\
$-\frac{2}{6}$ & $\frac{1}{3}$ & $-\frac{1}{6}$ & 0 & $\frac{1}{12}$ & $-\frac{1}{15}$ & 0 \\
$-\frac{3}{6}\left(-\frac{2}{4}\right)$ & $\pm \frac{1}{2}$ & 0 & $\pm \frac{1}{6}$ & 0 & $\pm \frac{1}{10}$ & 0 \\
$-\frac{4}{6}$ & $-\frac{1}{3}$ & $\frac{1}{6}$ & 0 & $-\frac{1}{12}$ & $\frac{1}{15}$ & 0 \\
$-\frac{3}{4}$ & $-\frac{1}{4}$ & $\pm \frac{1}{4}$ & $\frac{1}{12}$ & 0 & $-\frac{1}{20}$ & $\pm \frac{1}{12}$ \\
$-\frac{5}{6}$ & $-\frac{1}{6}$ & $-\frac{1}{6}$ & $\pm \frac{1}{6}$ & $\frac{1}{12}$ & $\frac{1}{30}$ & 0
\end{tabular}
\caption{ Ground state apex defect charges $q_A$ according to Eq.~\eqref{eq:qA_free} and \eqref{eq:ground_free} for hyperbolic cones commensurate with the triangular and
square lattices, in agreement with the numerics. Note that the apex charge of hyperbolic cones simply changes sign compared to conventional cones in Table~\ref{tab:free_cone}. }
\label{tab:hyperbolic_cone}
\end{table}

\subsection{Tangential boundary conditions}

\begin{figure}[t]
	\centering
	\subfloat[$\chi=0$]
	{\raisebox{0.5\height}{\includegraphics[width=.2\textwidth]{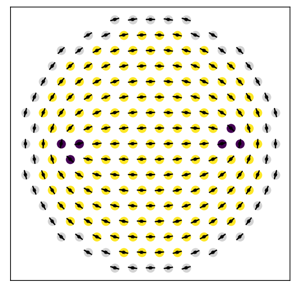}}}
	\hfill
 	\subfloat[$\chi=-1/3$]
	{\includegraphics[width=.39\textwidth]{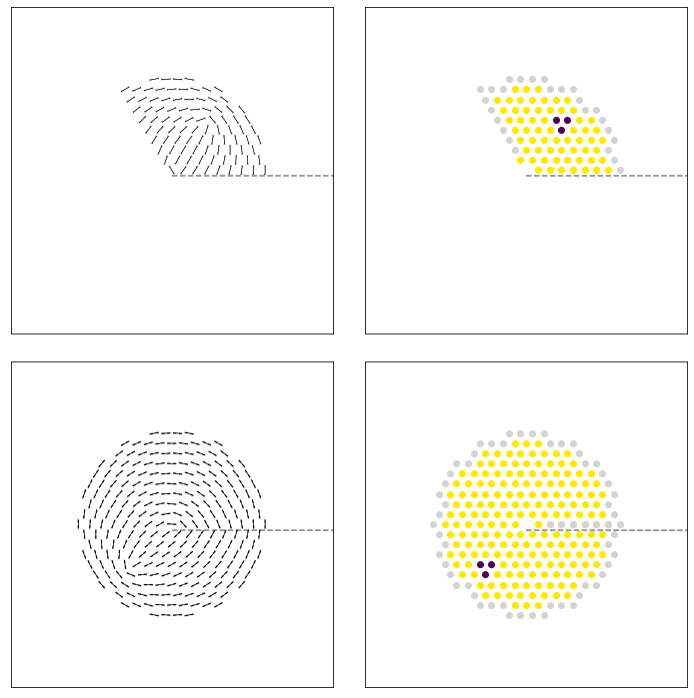}}
	\hfill
 	\subfloat[$\chi=-1/2$]
	{\includegraphics[width=.39\textwidth]{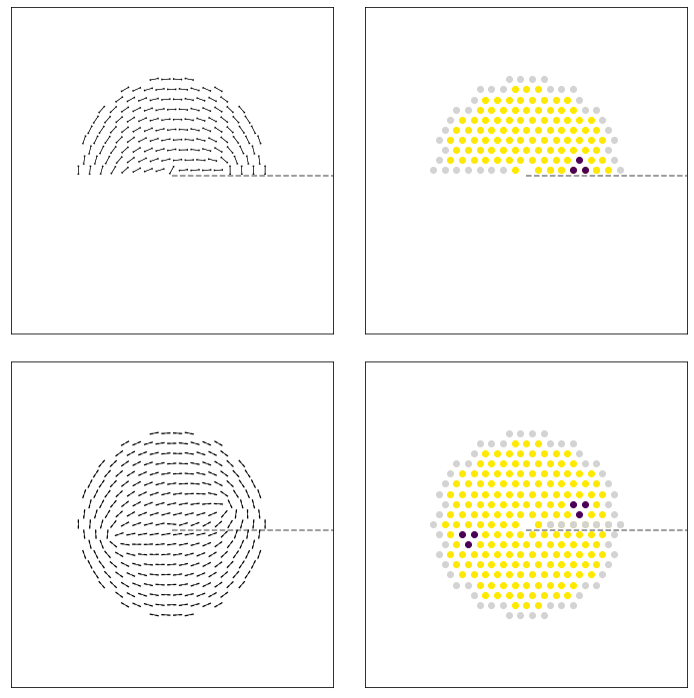}}
	\caption{Positive defect emission for a $p=2$ nematic liquid crystal on a hyperbolic cone as $\chi$ grows more negative, starting from a disk with $\chi=0$. For each value of $\chi$, the ground state texture obtained from numerical energy minimizations is shown, along with a colored plot or overlay indicating the location of the topological defects. The reader can refer to Fig.~\ref{fig:cone-construction}b for the construction and orientation of hyperbolic cones (consisting of a cut disk and an angular sector) in subfigures (b) and (c) in this figure. For $\chi=-1/3$, a sector with angle $2\pi/3$ has been inserted into a cut in a disk, while for $\chi=-1/2$, the inserted sector has angle $2\pi/2$, i.e., $180^\circ$. Note that the hyperbolic cone for $\chi=-1/3$ retains the two $+1/2$ flank defects of the disk, while for $\chi=-1/2$, a third $+1/2$ defect has appeared on the flanks. In each colored plot, sites constrained by either tangential boundary conditions or boundary conditions along the cut are indicated as gray, topological defects with charge $+1/2$ are blue, and remaining sites are yellow.}
	\label{fig:emission}
\end{figure}

\begin{figure}[t]
	\centering
	\includegraphics[width=\columnwidth]{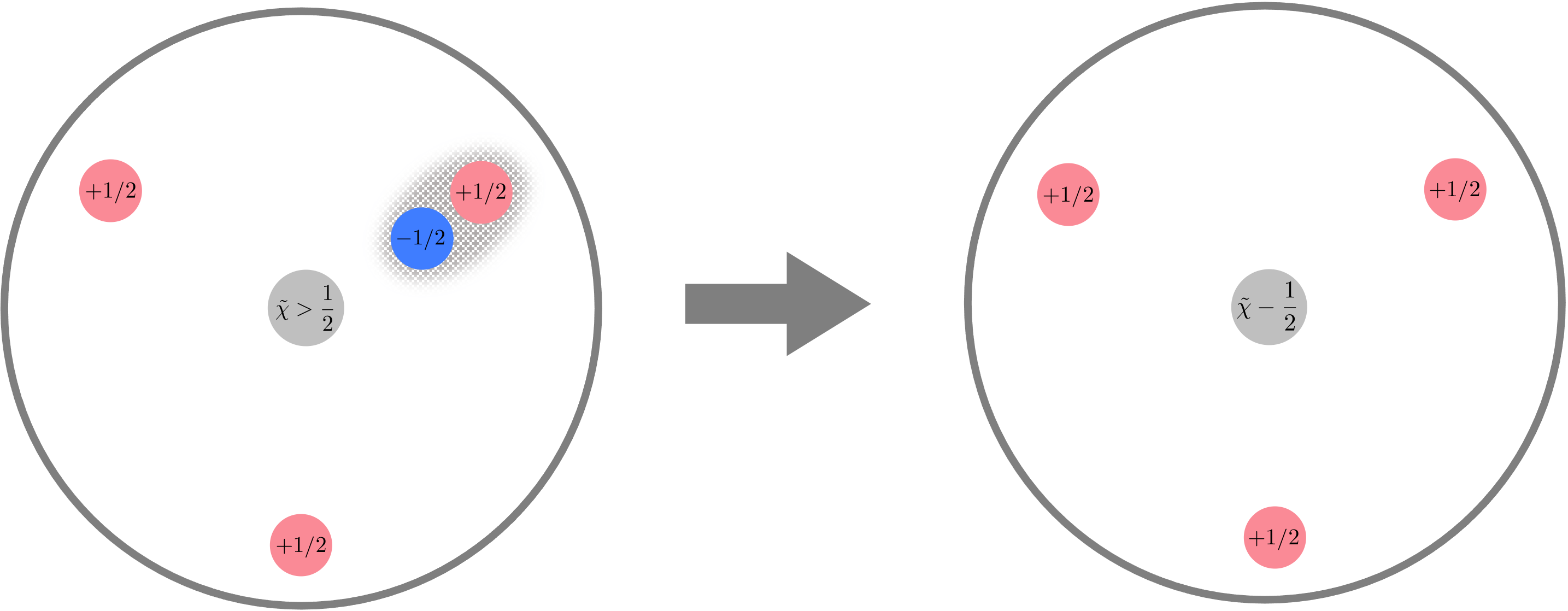}
	\caption{Mechanism by which the defect number of the flanks of a hyperbolic cone (shown here schematically in projection) can change for a $p=2$ nematic liquid crystal with tangential boundary conditions. Extra $+1/2$ flank defect relative to the two present on a disk originates from the unbinding of a $\pm 1/2$ defect dipole. The apex charge $\tilde \chi \equiv -\chi > 0$ changes from a positive value $\tilde\chi > 1/2$ to $\tilde\chi - 1/2$ when it absorbs the $-1/2$ part of the dipole.}
	\label{fig:unbinding}
\end{figure}

Similar to conventional cones, hyperbolic cones with tangential boundary conditions are more complex than those with free boundary conditions. In contrast to conventional cones, where the apex can absorb positive topological defects, here the apex charge is positive and the apex can in fact \emph{emit} positive topological defects! It can do this by nucleating a neutral pair of plus-minus defects. See Fig.~\ref{fig:emission} for the texture obtained from numerical simulations and Fig.~\ref{fig:unbinding} for a schematic of defect emission.

\begin{figure}[t]
	\centering
	\subfloat[]
	{\includegraphics[width=1.\columnwidth]{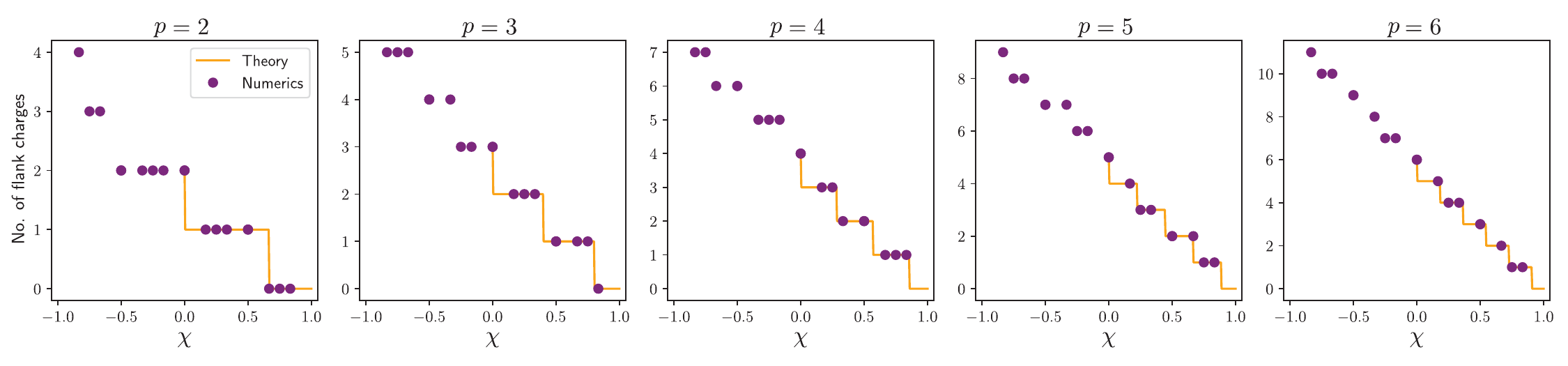}}
	\\
	\subfloat[]
	{\includegraphics[width=1.\columnwidth]{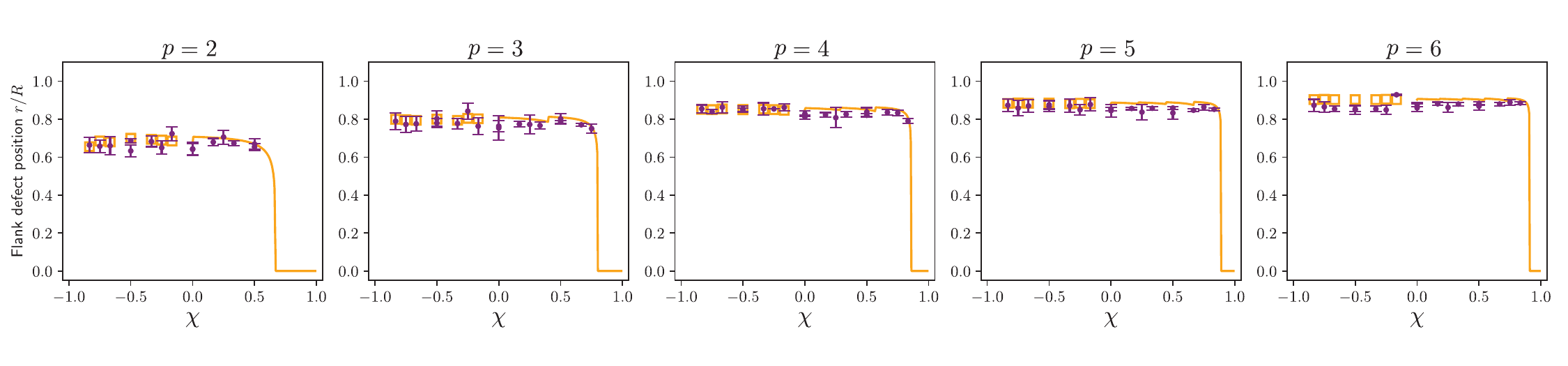}}
	\caption{Top row: plots of number of flank charges for both positive and negative $\chi$, thus extending the results of Fig.~\ref{fig:cone-agreement}. Purple dots are from numerical energy minimization, and orange lines for positive $\chi$ are the theoretical prediction for defect absorption transitions on conventional cones [Eq.~\eqref{eq:chi_c}]. Bottom row: plots of flank defect positions in the ground states of $p$-atics on hyperbolic cones as a function of $\chi$. Purple markers are from numerical energy minimization, and orange curves for $\chi > 0$ and orange squares for $\chi < 0$ are the theoretical predictions from Eq.~\eqref{eq:dxR}.}
	\label{fig:hyperbolic_cone-agreement}
\end{figure}

As shown in Fig.~\ref{fig:unbinding}, above a critical value of $\tilde\chi \equiv -\chi > 0$, the positive apex charge triggers the nucleation of a $\pm 1/2$ defect pair which separates, leaving behind three $+1/2$ defects on the hyperbolic cone flank, and an effective apex charge $\tilde\chi - 1/2$ at the apex. As shown in Fig.~\ref{fig:hyperbolic_cone-agreement}, the transition to three flank defects occurs when $\chi\approx-1/2$. 

\begin{figure}[t]
    \centering
    \includegraphics[width=.9\columnwidth]{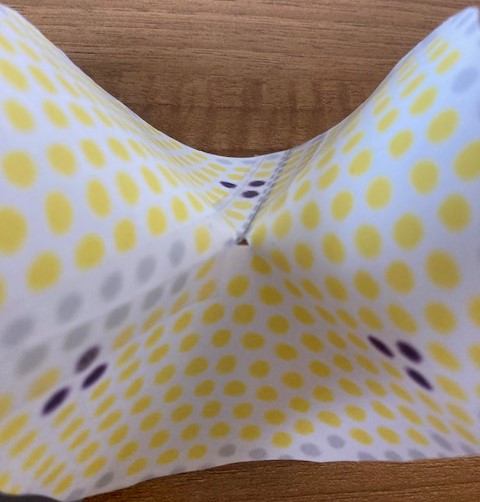}
    \caption{Locations of the three $+1/2$ defects surrounding the apex on a hyperbolic cone with $\chi=-1/2$, i.e., with a semicircular wedge added to a flat disk. The bending energy of the paper produces a characteristic hyperbolic shape in three dimensions. See Fig.~\ref{fig:emission}(c) for the corresponding director field
}
    \label{fig:hyperbolic_cone3D}
\end{figure}

See Fig.~\ref{fig:hyperbolic_cone3D} for a three-dimensional image of three $+1/2$ defects in the ground state for a hyperbolic cone for $\chi=-1/2$.

Although we lack at present a detailed theory for the transitions of defect emission, if we are given the total number of defects, then Eq.~\eqref{eq:dxR} predicts the defect positions remarkably well, see Fig.~\ref{fig:hyperbolic_cone-agreement}.

\section{Discussion}
\label{sec:discussion}

In this article, we have reviewed our understanding of liquid crystal ground states on cones, both conventional and hyperbolic. To summarize, topological defects act as point charges, and the cone apex behaves as if it were also charged. Thus, conventional cones tend to absorb positive defects at the apex, whereas hyperbolic cones can emit positive defects from the apex.

Much work remains to be done. Here, we have focused on the ground state configurations, but it is very natural to consider the non-equilibrium setting associated with active matter. Introducing activity in the context of nematics leads to geometric contributions to the motility of active topological defects~\cite{vafa2023active}. In additional work~\cite{vafa2023periodic}, preliminary investigations have revealed a rich phase diagram of $p=2$ allowed dynamical states, which exhibits not only single flank defect orbits and two flank defect orbits, but also transitions between them via defect absorption, defect emission, and defect pair creation via activity at the apex. 

In a related direction, it would also be worth studying the equilibrium statistical mechanics of defect configurations on cones with both tangential and free boundary conditions at finite temperatures. With increasing temperatures, entropic effects might also cause the cone apex to emit defects or possibly even alter defect unbinding transitions.

\acknowledgments{It is a pleasure to contribute to this volume honoring Uwe T\"{a}uber. This work is partially supported by the Center for Mathematical Sciences and Applications at Harvard University (F. V.), and by the Harvard Materials Research Science and Engineering Center via Grant DMR-2011754 (D.R.N.). We also acknowledge conversations with Cheng Long.}

\bibliography{refs}
	
\end{document}